\documentclass[epj, nopacs]{svjour}

\usepackage{booktabs}
\usepackage{graphicx}

\usepackage{xurl}

\usepackage[
    backend=biber,      
    style=numeric-comp, 
    sorting=none,       
    giveninits=true,    
    doi=true,           
    isbn=false,          
    eprint=false
]{biblatex}

\usepackage{ulem}
\usepackage{xurl}

\usepackage{xspace}
\usepackage{amsmath}
\usepackage[colorlinks=true, linkcolor=blue, citecolor=blue, urlcolor=blue]{hyperref}
\def\sharpy{\texttt{SHARPy}\xspace}

\usepackage{orcidlink}

\usepackage{academicons}   
\usepackage{xcolor}        

\begin{document}

\title{The Sequential Monte Carlo goes NUTS:\\
Boosting Gravitational-Wave Inference}

\author{Gabriele Demasi \thanks{\emph{E-mail:} gabriele.demasi@unifi.it} \inst{1,2}\orcidlink{0009-0009-5320-502X}
  \and Giulia Capurri \inst{3, 4}\orcidlink{0000-0003-0889-1015} \and Massimo Lenti\inst{1, 2}\orcidlink{0000-0002-2765-3955} \and Angelo Ricciardone \inst{3, 4}\orcidlink{0000-0002-5688-455X} \and Barbara Patricelli\inst{3, 4}\orcidlink{0000-0001-6709-0969} \and Adriano Frattale Mascioli\inst{5, 6}\orcidlink{0000-0002-0155-3833} \and Lorenzo Piccari\inst{5, 6}\orcidlink{0009-0000-0247-4339} \and Saulo Alberquerque\inst{7,2}\orcidlink{0000-0003-2911-9358} \and Gianluca M. Guidi\inst{7,2}\orcidlink{0000-0002-3061-9870} \and Francesco Pannarale\inst{5,6}\orcidlink{0000-0002-7537-3210} \and Giulia Stratta \inst{8,9}\orcidlink{0000-0003-1055-7980}\and Walter Del Pozzo\inst{3,4}\orcidlink{0000-0003-3978-2030}}
%
%
\institute{
Dipartimento di Fisica e Astronomia, Università degli Studi di Firenze,
Via Sansone 1, Sesto Fiorentino (Firenze) I-50019, Italy 
\and 
INFN, Sezione di Firenze, Sesto Fiorentino (Firenze) I-50019, Italy
\and 
Dipartimento di Fisica “Enrico Fermi”, Università di Pisa,
Largo Bruno Pontecorvo 3, Pisa I-56127, Italy
\and
INFN, Sezione di Pisa, Largo Bruno Pontecorvo 3, Pisa I-56127, Italy
\and
Dipartimento di Fisica, ``Sapienza'' Universit\`a di Roma, Piazzale Aldo Moro 5,  00185, Roma, Italy
\and
Dipartimento di Fisica, Sezione INFN Roma, Piazzale Aldo Moro 5,  00185, Roma, Italy
\and
Universit\`a degli Studi di Urbino ``Carlo Bo'', I-61029 Urbino, Italy
\and
INAF, Osservatorio di Astrofisica e Scienza dello Spazio, Via Piero Gobetti 101, 40129 Bologna, Italy
\and
INFN, Sezione di Bologna, viale Carlo Berti Pichat 6/2, 40127 Bologna, Italy}
\date{Received: date / Revised version: date}
%
\abstract{
 Sequential Monte Carlo (SMC) methods have recently been applied to gravitational-wave inference as a powerful alternative to standard sampling techniques, such as Nested Sampling. At the same time, gradient-based Markov Chain Monte Carlo algorithms, most notably the No-U-Turn Sampler (NUTS), provide an efficient way to explore high-dimensional parameter spaces. In this work we present \sharpy\ , a Bayesian inference framework that combines the parallelism and evidence-estimation capabilities of SMC with the state-of-the-art sampling performance of NUTS. Moreover, \sharpy \ exploits the local geometric structure of the posterior to further improve efficiency. Built on JAX, a high-performance computing framework that enables automatic differentiation and hardware acceleration, \sharpy\  performs gravitational-wave inference on binary black-hole events in around ten minutes, yielding posterior samples and Bayesian evidence estimates that are consistent with those obtained through Nested Sampling. This work sets a new milestone in Gravitational-Wave inference with likelihood-based methods and
paves the way for model comparison tasks to be accomplished in minutes.
%
} 
\maketitle

\section{Introduction}

\noindent

Once a gravitational-wave (GW) signal from a compact binary coalescence is detected by the LIGO-Virgo-KAGRA (LVK) Collaboration detectors \cite{2015:LIGO,2015:VIRGO,2013:KAGRA}, Bayesian inference is the tool used to understand the properties of the source that generated it and to determine which of the available models describes better the data \cite{2019:Thrane_Talbot}. 
The inference outcome forms the baseline for analyses aimed at, for instance, inferring cosmological parameters, testing General Relativity and constraining the population properties of binary black holes (BBH) and binary neutron stars in the Universe \cite{2025:O4_population,2025:O4_cosmo,2021:O3_TGR}. Nested Sampling \cite{2006:Skilling,2022:Ashton_NS} is one of the standard algorithms for GW inference used by the LVK Collaboration \cite{2025:O4_methods}.
Despite being known for its robustness, Nested Sampling is computationally demanding and intrinsically sequential, requiring inference runs that may extend to hours or even days. 
Sequential Monte Carlo (SMC) methods \cite{2006:Doucet_SMC}, while known for a long time, have only recently been applied to GW astronomy as an alternative to Nested Sampling for parameter estimation and model comparison \cite{2022:Karamanis_Preconditioned,2025:Williams_Validating_SMC, 2025:ASPIRE}. Moreover, SMC can be reformulated to operate as a Nested Sampler~\cite{Salomone2025Unbiased}. 
At their core, SMC algorithms evolve a population of particles from a reference distribution, often the prior, to a target distribution (the posterior) through a temperature ladder scheme. By means of a repeated use of importance sampling, this scheme allows for an unbiased estimation of the evidence, offering a compelling alternative to Nested Sampling.
 For the exploration of the parameter space, however, SMC algorithms still rely on a transition Markov kernel, such as a Markov Chain Monte Carlo (MCMC); this acts as a performance bottleneck. On the other hand, gradient-based kernels, such as the Hamiltonian Monte Carlo (HMC) \cite{DUANE1987216, 2011:Neal_HMC, Porter:2013wwa, 2025:DEEP_HMC}, use gradient information to partially suppress the typical random walk behavior of the MCMC kernel and have started to be used in the SMC scheme\cite{varsi2024general, millard2025incorporating,devlin2021no}. The No-U-Turn-Sampler, an improved version of the HMC that addresses some of its limitations, has emerged as a prominent tool for sampling in high-dimensional space \cite{2011:NUTS}, but it has never been applied to single-event GW inference.

Because of their intrinsic parallelism, SMC algorithms are naturally well suited for modern hardware accelerators such as GPUs, which enable massive speed-ups.

In this work, we present \sharpy, a framework for accelerating GW inference that leverages on the combination of the intrinsic parallelism of SMC, the use of the efficient parameter-space exploration of the NUTS and the implementation in \texttt{JAX} \cite{jax2018github}, a computing framework that allows for automatic differentiation and hardware-accelerated computation.
This combination yields a significant acceleration compared to standard parameter estimation algorithms, while preserving accuracy and, to the best of our knowledge, has not been exploited in previous SMC implementations.

We show that \sharpy produces high-quality posterior samples and precise evidence estimates for both simulated and real GW data, achieving performance comparable to state-of-the-art samplers while requiring only a small fraction of the computational time. 

The paper is organized as follows: in section \ref{se:GW_Inference} we state the general problem of GW Bayesian inference; in section \ref{sec:SMC} we give an introduction to the SMC method; in section \ref{sec:NUTS} we describe the HMC and the NUTS; in section \ref{sec:SHARPy} we explain the main features of \sharpy while in section \ref{sec:GW_inf} we show its performance when applied to both real and simulated data, making a systematic comparison with other samplers. Finally, we draw our conclusions in section \ref{sec:Conclusions}.

\section{Gravitational Wave Bayesian Inference}\label{se:GW_Inference}
\noindent
Bayesian inference revolves around Bayes’ theorem, which provides a framework for updating our knowledge of the model parameters in light of observed data. Given a hypothesis or model $H$, the theorem states that the posterior probability distribution of the parameters $\theta$ conditioned on the data $d$ is
\begin{equation}
    \label{th:Bayes}
    p(\boldsymbol{\theta}|d, H) = \frac{\mathcal{L}(d|\boldsymbol{\theta}, H) \mathcal{\pi}(\boldsymbol{\theta}| H)}{p(d| H)}, 
\end{equation}
where the likelihood $ \mathcal{L}(d|\boldsymbol{\theta}, H)$ describes how likely it is to observe the data $d$ for given parameter values $\boldsymbol{\theta}$,
the prior $\pi(\boldsymbol{\theta}| H)$ reflects our knowledge of the parameters $\boldsymbol{\theta}$ before observing the data 
and $\mathcal{Z} = p(d| H)  = \int d\boldsymbol{\theta}\mathcal{L}(d|\boldsymbol{\theta}, H)\pi(\boldsymbol{\theta}|H)$ is the Bayesian evidence, which ensures proper normalization of the posterior. The evidence can be ignored if we are interested in estimating the parameters $\boldsymbol{\theta}$ while it plays a central role in model comparison, as it quantifies how well the model $H$ explains the observed data after integrating over all possible parameter values.

 Writing the data as $d = n + h(\boldsymbol{\theta})$, with $n$ the detector noise and $h(\boldsymbol{\theta})$ the GW signal model, and assuming the noise to be stationary and Gaussian, the frequency-domain log-likelihood takes the form \cite{Christensen:2022bxb}:
\begin{equation} \label{eq_loglikelihood}
    \log\mathcal{L}(d|\boldsymbol{\theta}) = -\frac{1}{2}\left\langle d-h(\boldsymbol{\theta})|d - h(\boldsymbol{\theta})\right\rangle, 
\end{equation}
with $ \left\langle a|b\right\rangle = 4\mathrm{Re} \int_0^\infty \frac{a^{*}(f)b(f) }{S_{n}(f)}df$, where $S_{n}(f)$ is the one-sided power spectral density (PSD) of the detector noise \cite{LIGOScientific:2019hgc}. 
For further details on Bayesian inference in GW astronomy, see \cite{Christensen:2022bxb}.
A typical compact binary coalescence signal is described by roughly fifteen parameters. For a BBH event, these include eight intrinsic parameters related to the component masses and spins, and seven extrinsic parameters, such as the luminosity distance, inclination, sky location (right ascension and declination), coalescence phase and time, and polarization.

Stochastic sampler methods, such as MCMC or Nested Sampling, are typically employed to explore this high-dimensional parameter space and perform Bayesian inference \cite{2015:LAL_inference, 2018:ashton_bilby}, although other strategies are also employed\cite{Lange:2018pyp}. Alternative schemes based on simulation-based inference have begun to be adopted \cite{2021:Dax_DINGO, 2024:De_Santi, Gabbard:2019rde, Chua:2019wwt}.

In general, the inference computational cost is primarily driven by two main factors:
\begin{enumerate}
    \item the dimensionality of the parameter space, since the efficiency of proposing new points decreases as the number of parameters grows, and
    \item the cost of waveform generation, which scales with the number of frequency bins used to evaluate the waveform.
\end{enumerate}
The latter depends on both the sampling rate and the duration of the signal. In order to speed up the inference of GW signals, it is possible to act on both aspects. Starting from the cost of waveform generation, it is possible to exploit GPU acceleration \cite{2023:Wong_JIM, 2024:Robust_Wouters, Prathaban:2025qgg}, that allows for the parallel evaluation of waveforms, and to have strategies for reducing the number of frequency points with Reduced Order Quadrature \cite{2020:ROQ, Smith:2016qas}, Multibanding \cite{2021:Multibanding_MOrisaki} or  Relative Binning \cite{2023:Relative_Binning, Cornish:2010kf}. 
Regarding proposal efficiency, several strategies have been explored in the literature. Some approaches employ Normalizing Flows to construct data-driven proposal distributions \cite{2021:Williams_Nessai}, others use gradient-based methods to more effectively traverse the parameter space \cite{2023:Wong_JIM, 2025:DEEP_HMC} and some rely on ad-hoc reparameterizations tailored to the structure of the problem \cite{2022:Cogwheel}.

In this work, as described in the following sections, we adopt  gradient-based proposals  mechanisms within a Sequential Monte Carlo framework, leveraging GPU acceleration to achieve an efficient and scalable approach to GW inference. 

\begin{figure}
\resizebox{0.5\textwidth}{!}{%
  \includegraphics{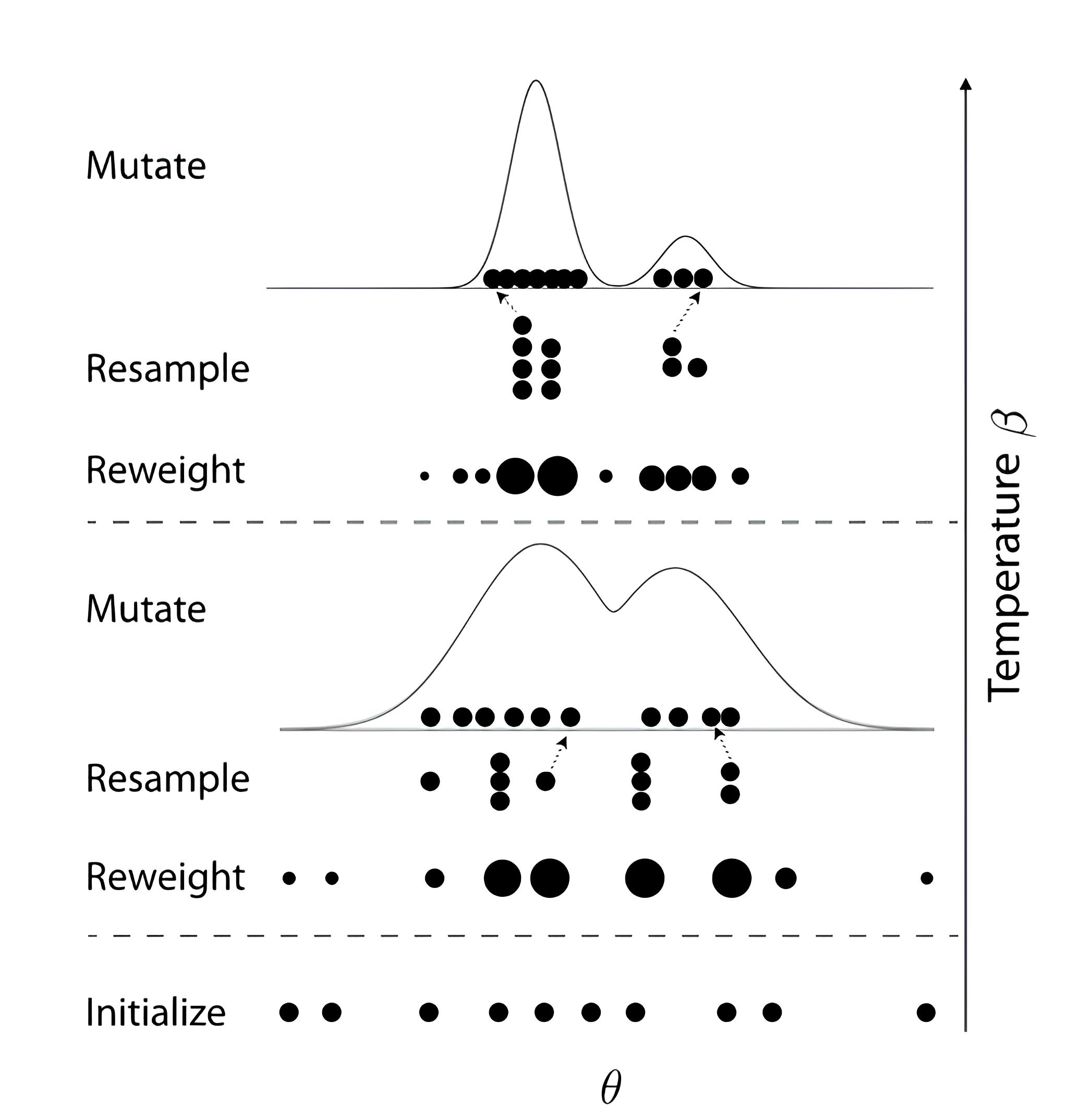}
}
\caption{Illustration of the SMC algorithm applied on a bimodal Gaussian mixture distribution. Particles are first randomly drawn from the prior. Then they are reweighed according to tempered distribution with a certain $\beta$ and resampled according to these weights so that particles the lie in high likelihood regions are selected. At the end of each SMC iteration, in the mutation step, particles explore the space with some transition kernel. Adapted from \cite{hinne2025introduction}.}
 \label{fig:smc_picture}
 \end{figure}

 \section{Sequential Monte Carlo}\label{sec:SMC}
We use Sequential Monte Carlo (SMC) as an alternative strategy to perform GW inference. SMC was originally introduced as a framework for analyzing time-series models in which data arrive sequentially \cite{2001:SMC_introduction}. More recently, it has been successfully extended to static inference problems using temperature-annealing schemes, in which the sampler begins with an easier, smoothed version of the target distribution and gradually evolves toward the true posterior. 
This allows us to mitigate multimodal parameter space issues and alleviate convergence issues that often challenge traditional MCMC methods \cite{2022:SMC_invitation}.

In this framework, a population of $N_\mathrm{P}$ particles is initially sampled from the prior distribution. These particles are then evolved through a sequence of intermediate distributions that are built to gradually increase the influence of the likelihood on the target distribution, so that the final distribution matches the posterior.
Formally, the sequence is constructed introducing a temperature parameter $\beta_t$ at each SMC iteration $t$:
\begin{equation}
    p_t(\boldsymbol{\theta}|d) = \frac{\mathcal{L}(d|\boldsymbol{\theta})^{\beta_t}\pi(\boldsymbol{\theta)}}{\mathcal{Z}_t}, 
\end{equation}
where, in comparison to Eq.\,\eqref{th:Bayes} we suppressed $H$ to achieve a lighter notations, and $\mathcal{Z}_t = \int d\boldsymbol{\theta}\mathcal{L}(d|\boldsymbol{\theta})^{\beta_t}\pi(\boldsymbol{\theta})$.  The SMC starts with $\beta_0 = 0$ and $p_0(\boldsymbol{\theta}|d) \equiv \pi(\boldsymbol{\theta)}$, which is the prior, and it finishes after $T$ iterations when $\beta_T = 1$ and $p_T(\boldsymbol{\theta}|d)$ is the full posterior.
Each iteration consists of the three steps described below, which are also schematically depicted in Fig.\,\ref{fig:smc_picture}.
\begin{enumerate}
    \item \underline{\textbf{Reweighting}}\vspace{0.01\linewidth}\\
    At iteration $t-1$, each particle $i$ is assigned a weight that determines its ``fitness'' to the distribution at iteration $t$:

\begin{equation}
        w^{(i)}_t = \frac{p_t(\boldsymbol{\theta}_{t-1}|d)}{p_{t-1}(\boldsymbol{\theta}_{t-1}|d)} = \mathcal{L}(\boldsymbol{\theta}_{t-1}|d)^{\beta_t - \beta_{t-1}}. 
    \end{equation}
    If the distributions at iteration $t$ and $t-1$ are very far from each others, weights tend to become uneven, with the majority of particles having low weights at the expense of few with very high ones. 
    The effective sample size at iteration $t$, defined as \cite{Ashton:2025xba}
    \begin{equation}
        \mathrm{ESS}_t = \frac{\left(\sum_{i=1}^Nw^{(i)}_t\right)^2}{\sum_{i=1}^N \left(w^{(i)}_t\right)^2}, 
    \end{equation}
    provides a quantitative measurement of the weight degeneracy within the population of particles. 
    The temperature is  determined adaptively during the run. At each step, $\beta_t$ is computed by solving the following equation:
    \begin{equation}
        \textrm{ESS}\left(\beta_t\right) - \alpha N_P = 0,
    \end{equation}
    where $\alpha \in (0,1]$.
    In practice, we choose the next $\beta_t$ by requiring the ESS remains constant throughout the run, introducing a parameter $\alpha$ that controls the fraction of effective particles with respect to the total number of particles $N_P$.
    \item \underline{\textbf{Resampling}}\vspace{0.01\linewidth}\\
    At this step, particles are resampled according to their normalized weights, usually with multinomial resampling. This allows to replace ``low-weight'' particles (those lying in low-likelihood regions) with higher weight particles. As the temperature $\beta_t$ increases, the influence of the likelihood on the target distribution grows and the features become more highlighted. The reweighting procedure ensures the particle are correctly weighted according to the emerging features.
    \item \underline{\textbf{Mutation}}\vspace{0.01\linewidth}\\
    Finally, each particle is mutated with a transition kernel, typically an MCMC. This step is fundamental to prevent particle degeneracy and to ensure a good coverage of the 
    parameter space, which is particularly important for accurate evidence estimation. A key feature of SMC is that, at each iteration, each particle is evolved
    independently from each other. Since the mutation step is the most computationally expensive one, this independence enables massive parallelization and significantly reduces overall wall-clock time. 
    The more the transition kernel moves are efficient in exploring the parameter space,  the more the moved particles follow the target distribution, resulting in high values of the ESS and an overall reduction on the number of SMC iterations. 
\end{enumerate}

\subsection{Evidence computation}\label{sec:evidence}
    After each reweighting step the ratio of normalizing constants $\frac{\mathcal{Z}_t}{\mathcal{Z}_{t-1}}$ can be computed as: 
    \begin{equation}\label{eq:evidence}
        \frac{\mathcal{Z}_t}{\mathcal{Z}_{t-1}} = \frac{1}{N}\sum_{i= 1}^{N}  w^{(i)}_t.
    \end{equation}
    Assuming that the prior is normalized and hence $\mathcal{Z}_0 $ = 1, at the end of the last iteration the full evidence $\mathcal{Z}$ can be estimated as the product of the evidence ratios at each SMC iteration \cite{chopin2020introduction}:
    \begin{equation}
        \mathcal{Z} = \frac{\mathcal{Z}_T}{\mathcal{Z}_{T-1}}\times ... \times \frac{\mathcal{Z}_1}{\mathcal{Z}_{0}}.
    \end{equation}

\section{No-U-Turn-Sampler}\label{sec:NUTS}
As highlighted in the previous discussion, a good transition kernel is fundamental to maintain a high ESS within a low number of SMC iterations. Stochastic samplers, such as MCMC, propose random jumps in the parameter space, resulting in a low efficiency in the exploration. Gradient-based methods such as Hamiltonian Monte Carlo (HMC) suppress the MCMC random behavior by evolving the position in the parameter space based on Hamiltonian dynamics \cite{2011:Neal_HMC}. 
In its standard implementation, the parameter space $\boldsymbol{\theta}$ is augmented with auxiliary momentum variables $\boldsymbol{\mathrm{r}}$. 
The momentum variables are drawn from a multivariate normal distribution, $\boldsymbol{\mathrm{r}} \sim \mathcal{N}(0, M)$, with the covariance matrix $M$ that can be seen as a metric defined on the parameter space. It is therefore possible to define the Hamiltonian $
\mathcal{H}(\boldsymbol{\theta}, \boldsymbol{\mathrm{r}}) = \  U(\boldsymbol{\theta})+ K(\boldsymbol{\mathrm{r}}) 
    = -p(\boldsymbol{\theta|d)} + \dfrac{1}{2}\boldsymbol{r}M^{-1}\boldsymbol{r}$, 
where the first (second) term represents the potential (kinetic) energy. 
For proposing a new point, the system is evolved using Hamilton's equations:
\begin{align}
\frac{d\boldsymbol{\theta}}{dt} &= \frac{\partial H}{\partial \mathbf{r}} = \nabla_{\boldsymbol{r}}K(\boldsymbol{{r}}) \,,\\
\frac{d\mathbf{r}}{dt} &= -\frac{\partial H}{\partial \boldsymbol{\theta}} = -\nabla_{\boldsymbol{\theta}} U(\boldsymbol{\theta})\,.\label{eq:ham2}
\end{align}

We solve Hamilton's equations using the explicit leapfrog integrator \cite{2011:HMC_NEAL}. 
Each leapfrog step updates the position $\boldsymbol{\theta}$ and $\boldsymbol{r}$ as follows: 

\begin{align}
\boldsymbol{r}\left(t+\frac{\epsilon}{2}\right)
&= \boldsymbol{r}(t) - \frac{\epsilon}{2}\frac{\partial U}{\partial \boldsymbol{\boldsymbol{\theta}}}\bigl(\boldsymbol{\theta}\bigr), \\
\boldsymbol{\theta}(t+\epsilon)
&= \boldsymbol{\theta}(t) + \epsilon\,\frac{\partial K}{\partial \boldsymbol{r}}\Bigl(\boldsymbol{r}\left(t+\frac{\epsilon}{2}\right)\Bigr), \\
\boldsymbol{r}(t+\epsilon)
&= \boldsymbol{r}\!\left(t+\frac{\epsilon}{2}\right)
 - \frac{\epsilon}{2}\frac{\partial U}{\partial \boldsymbol{\theta}}\bigl(\boldsymbol{\theta}(t+\epsilon)\bigr),
\end{align}

where $\epsilon$ is the discretized time step. 


The system is evolved simulating Hamiltonian dynamics for $L$ leapfrog steps and,
after the evolution, the new point is accepted or rejected using the usual Metropolis-Hasting rule. 
The number of leapfrog steps $L$ is a difficult parameter to tune: if $L$ is too small the algorithm behaves like a stochastic sampler with the trajectories resembling a random walk. If $L$ is too large, being the dynamics approximately conservative, the the trajectory will eventually begin to retrace itself, wasting computation. Reference \cite{NUTS-2014} proposed the No-U-Turn Sampler (NUTS) to address this issue. The NUTS is an extension of the HMC that tunes automatically the integration length $L$. It checks when the Hamiltonian trajectory starts retracing its steps and makes a ``U-Turn''. This results in a well-tuned HMC without the need to tune it based on the specific problem, assuring an optimal exploration of the parameter space. For more details on the algorithm we refer to \cite{NUTS-2014}.

Integrating Hamilton's equations requires computing the gradient of the target distribution. For a long time, this has been the main bottleneck of the application of gradient-based methods to non-trivial and expensive target distributions, such as the ones in GW inference.Ref. \cite{2025:DEEP_HMC} proposed a solution based on Deep Neural Network. 
However, we adopted the auto-differentiation technique for computing derivatives, that has recently gained popularity.
It exploits the fact that every function in a computer is ultimately composed by a sequence of elementary operations and applies the chain rule systematically to compute exact derivatives, up to machine precision.

\begin{figure*}[h!!]
    \centering
    \begin{minipage}{0.49\linewidth}
    \resizebox{1\textwidth}{!}{%
  \includegraphics{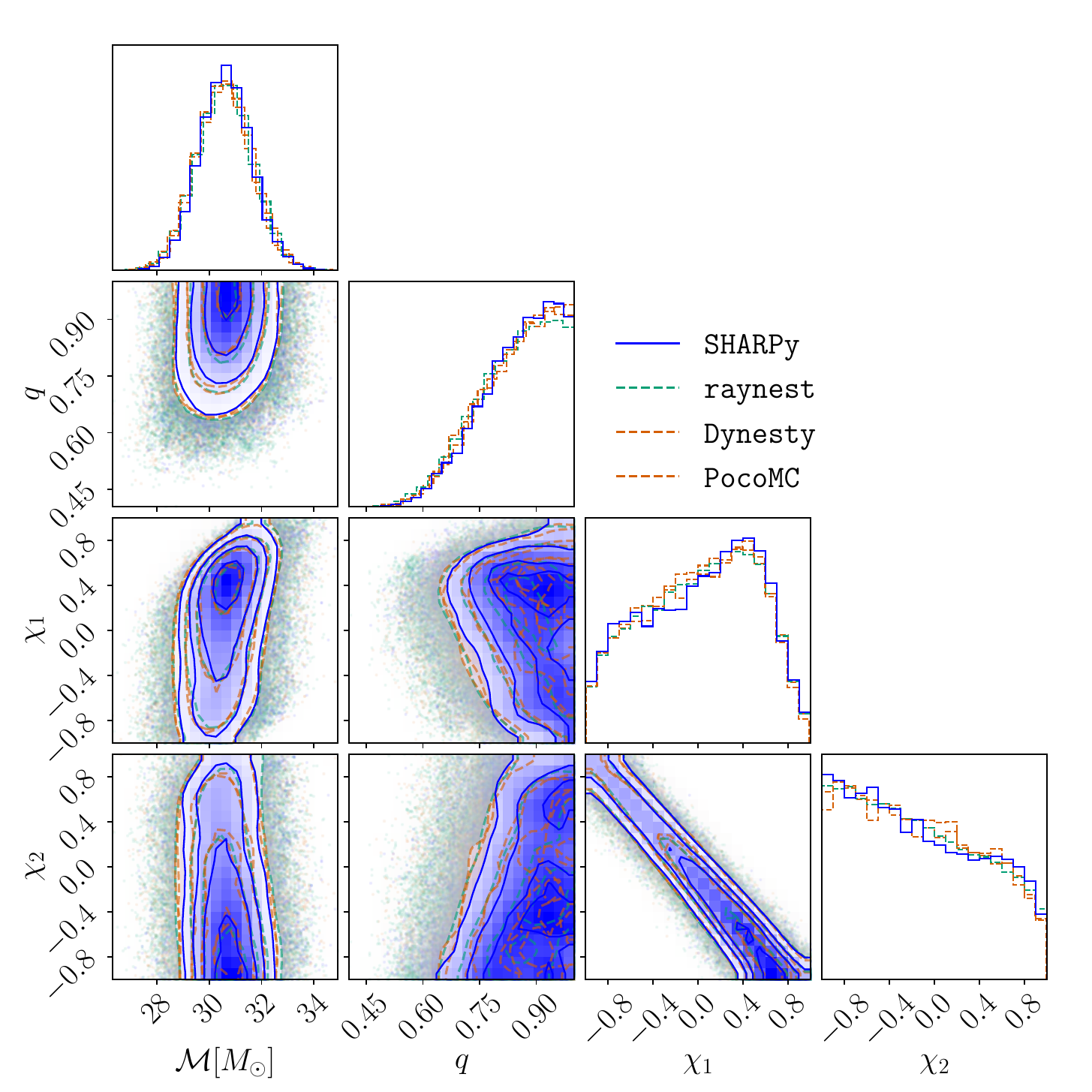}}
    \end{minipage}
    \hfill
    \begin{minipage}{0.49\linewidth}
    \resizebox{1\textwidth}{!}{%
  \includegraphics{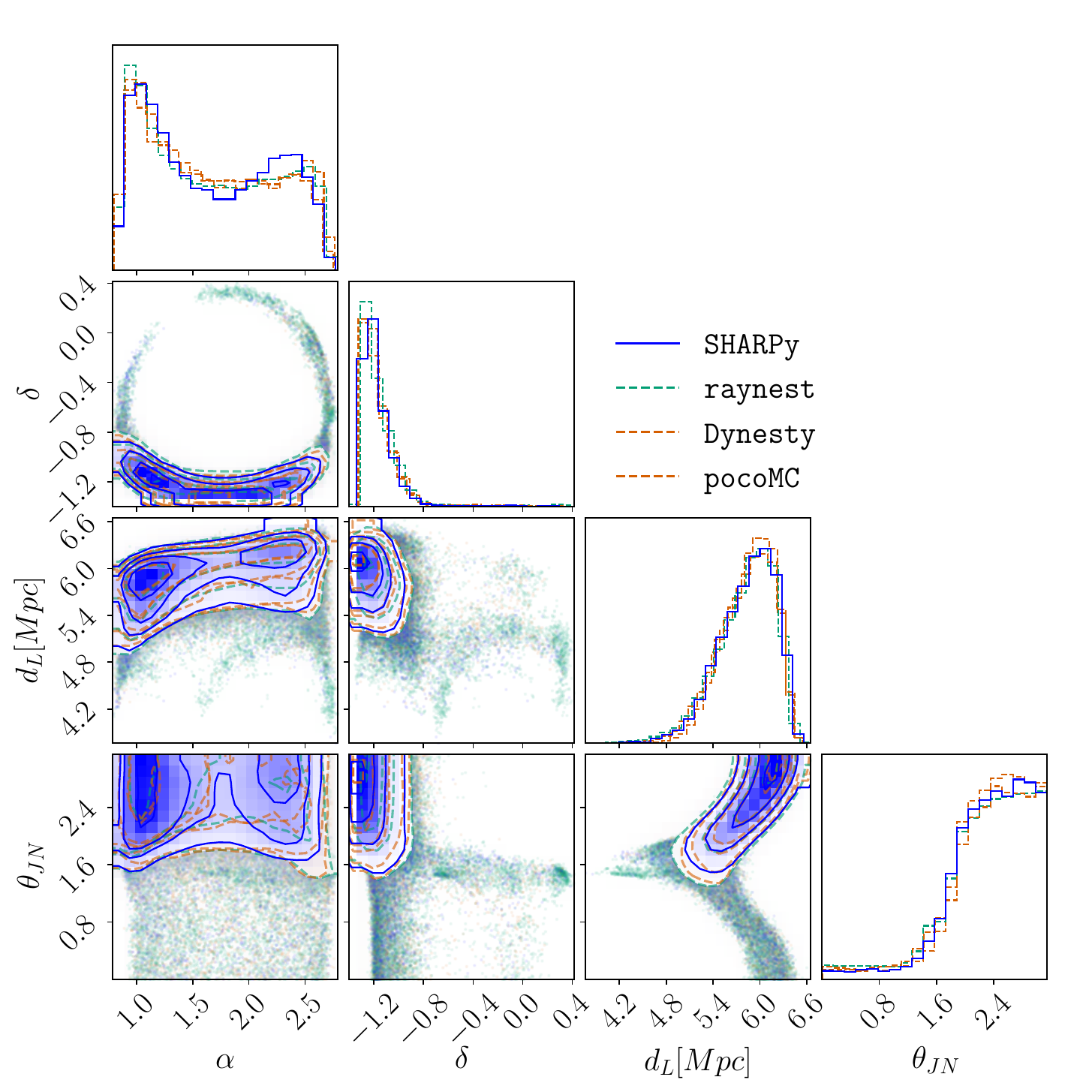}}
    \end{minipage}
    \caption{
    Comparison between the posterior samples of GW1509014 obtained with \sharpy (in blue, solid line) and the other reference samplers (dashed lines). 
    The corner plot on the left is limited to four intrinsic parameters (the chirp mass, the mass ratio and the two spin magnitudes), while the on the right shows four extrinsic parameters, namely the right ascension, the declination, the luminosity distance and the inclination angle between the line-of-sight and the total angular momentum.}
    \label{fig:samples_comparison}
\end{figure*}

\section{\sharpy}\label{sec:SHARPy}
\noindent
In this work we present \sharpy\footnote{Sequential Hamiltonian Riemann monte-carlo Python sampler}, a tool for gravitational-wave inference that integrates Sequential Monte Carlo with the No-U-Turn Sampler as a transition kernel. 
It has the following key features:
\begin{itemize}
    \item[] \underline{\textbf{Exploitation of local geometry}}\\
    As discussed in the previous section, the mass matrix $M$ can be seen as the metric of the parameter space \cite{2011:Riemann_Girolami}.
    In the most common implementations of HMC and NUTS, a global metric is fixed, taking it to be proportional to the identity or the Fisher Matrix \cite{2011:Neal_HMC, 2025:DEEP_HMC}. However, this approach fails to capture the local complexity of the posterior, therefore reducing the efficiency of the exploration of the parameter space. 
    To solve this problem, the Riemann Manifold Hamiltonian Monte Carlo was introduced \cite{2011:Riemann_Girolami}; it uses a position dependent metric. With this approach, however, Hamilton's equations are no longer separable  and they must be solved using implicit methods that are in general much more computationally intensive,leading to non-trivial and expensive integration techniques.

    In this work, we adopt a hybrid approach: 
    we use a position-dependent metric only at the beginning of each SMC iteration, while we fix it throughout the mutation step, in order to have a fixed matrix and therefore separable Hamilton's equations during the integration. In particular, at each SMC iteration and for each particle in the ensemble, we set the mass matrix $M$ to be the Hessian of the posterior 
    defined as:
    \begin{equation}
        \mathrm{H}_{ij}:=  \frac{\partial^2 p(\boldsymbol{\theta|}d)}{\partial\theta_i\partial\theta_j}.
    \end{equation}
    This way, the generation of momentum variables $\boldsymbol{\mathrm{r}}$ takes into account the local geometry of the distribution, enhancing the exploration of the parameter space. 
    \item[] \underline{\textbf{Boundary conditions}}\\
    Often, the parameter space in which the sampling happens is bounded. For example, in GW inference, the mass ratio is physically bounded to be in $(0, 1]$. Therefore, we implemented a strategy for dealing with HMC trajectories that go out of bounds, following the strategy adopted in \cite{2025:DEEP_HMC}. For certain parameters, such as the mass ratio,  we impose the boundary condition to be reflective, so that the particle exceeding the lower (upper) bound $\boldsymbol{\theta}^{\textrm{b}}_{\textrm{L}}$ ($\boldsymbol{\theta}^{\textrm{b}}_{\textrm{U}}$)  is elastically reflected:
    \begin{align}
      \boldsymbol{\theta} &\xrightarrow{} 2 \boldsymbol{\theta}^{\textrm{b}}_{\textrm{U,L}} - \boldsymbol{\theta}; \\
      \boldsymbol{\mathrm{r}} &\xrightarrow{} -  \boldsymbol{\mathrm{r}}
    \end{align}
    In the case of angular variables, instead, we impose periodic boundary conditions, such that the trajectory leaving the parameter space from one end re-enters from the opposite end:
      \begin{align}
      \boldsymbol{\theta} &\xrightarrow{}\boldsymbol{\theta}  -  (\pm\boldsymbol{\theta}^{\textrm{b}}_{\textrm{L}} -\mp \boldsymbol{\theta}^{\textrm{b}}_{\textrm{U}})\\
      \boldsymbol{\mathrm{r}} &\xrightarrow{}  \boldsymbol{\mathrm{r}}.
    \end{align}
    \item[] \underline{\textbf{Samples recycling}}\\
    The classic SMC scheme uses only particles from the last iteration to approximate the target distribution, wasting all particles from previous iterations. 
    In this work, following previous literature \cite{recycling:Nguyen, cornuet2012adaptive, 2024_Persistent},  we consider a pool of particles that includes also the ones from all iterations, as they are drawn from the following distribution: 
        \begin{equation}
        \tilde{p}(\boldsymbol{\theta}|d) = \frac{1}{T}\sum_{t=1}^{T}p_t(\boldsymbol{\theta}|d),
    \end{equation}
    where $p_t(\boldsymbol{\theta}|d)$ is normalized using the evidence estimated at each iteration. 
    Lastly,  we perform rejection sampling to 
    obtain independent and identically distributed (i.i.d.) samples from the target distribution.


    \item[] \underline{\textbf{\texttt{JAX} implementation}}\\
    \sharpy is entirely developed in \texttt{JAX}. The implementation of the algorithm is publicly available at
    \url{https://github.com/gabrieledemasi/sharpy}, while for the NUTS sampler and the waveforms we leverage respectively on  \texttt{BLACKJAX} \cite{2024:Blackjax} and \texttt{ripple} \cite{2023:Ripple}.
    The benefits that \texttt{JAX} brings to our purposes are twofold.
    First, it provides automatic differentiation, allowing for the computation of derivatives, specifically the   gradients and the Hessian needed in our case, through the repeated application of the chain rule to the function that needs to be differentiated, without the need of resorting to finite differences or symbolic derivation, that are either slow or inaccurate. 
    Second, because \texttt{JAX} is device-agnostic it enables running \sharpy on hardware accelerators such as GPUs.
    Additionally, \texttt{JAX} provides native support for just-in-time (\texttt{JIT}) compilation and automatic vectorization that, combined with hardware accelerators, makes it possible to effectively exploit the parallelism offered by the SMC scheme, resulting in a very low sampling wall-clock time as illustrated in the following sections.


\end{itemize}

\begin{figure}
    \centering
    \resizebox{0.5\textwidth}{!}{\includegraphics{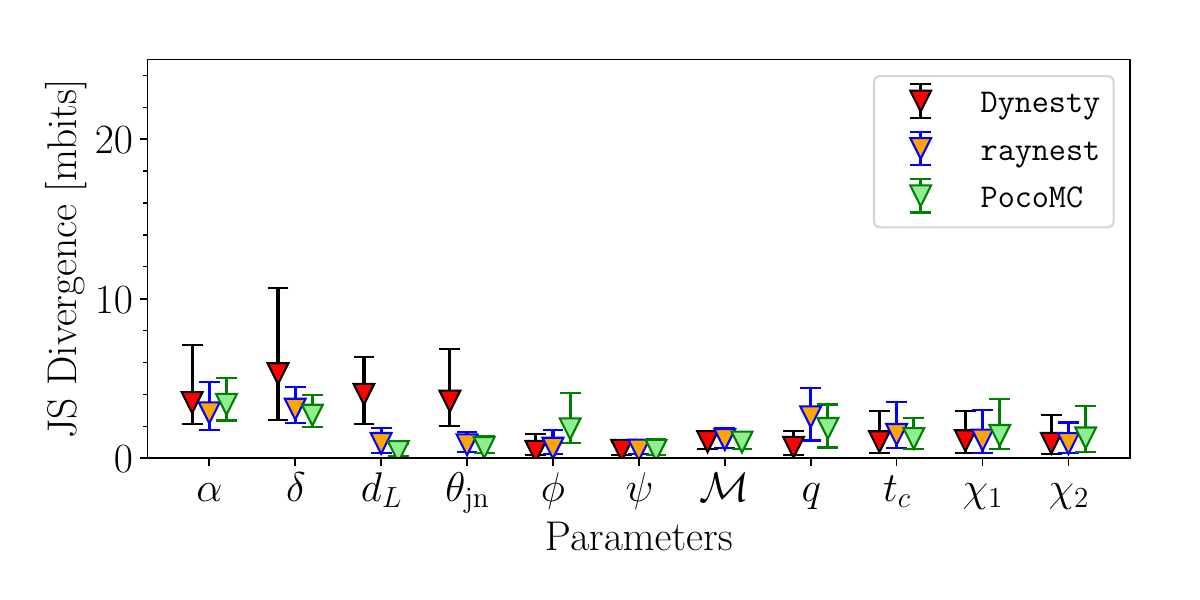}}
    \caption{Jensen-Shannon divergence, expressed in mbits, between the samples obtained with \sharpy and those obtained with the reference samplers in the GW150914 case. The triangles and the errobars indicate respectively the median and the 90\% credible intervals obtained from 100 indipendent runs with \sharpy.}
    \label{fig:js_divergence}
\end{figure}

\begin{figure}[!t]
    \centering
    \resizebox{0.5\textwidth}{!}{\includegraphics{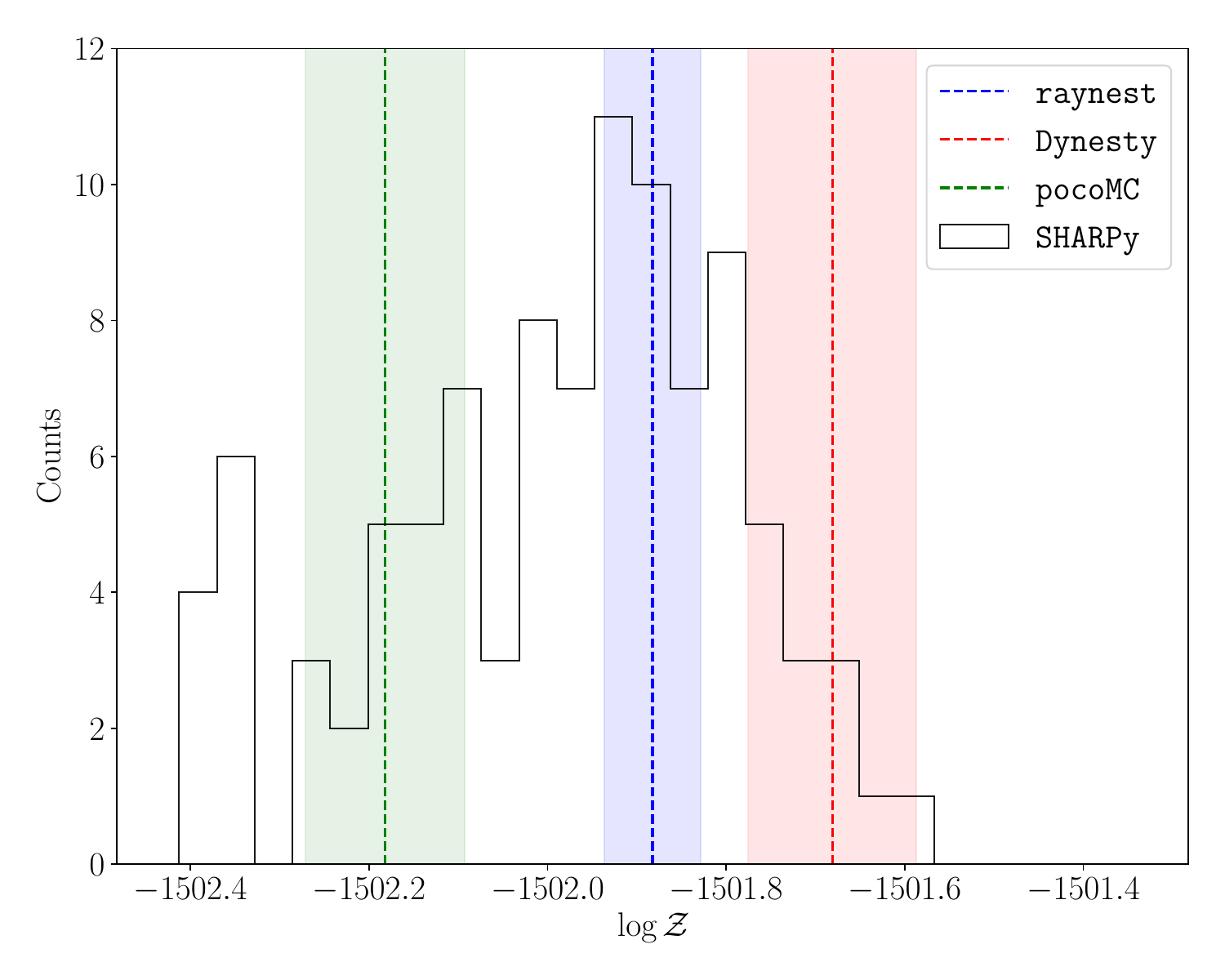}
    }\caption{
    Comparison between the evidence values obtained with 100 independent \sharpy runs on GW150914 data (solid line) and the evidence computed with \texttt{Dynesty}, \texttt{raynest} and \texttt{pocoMC} (dashed lines). The shaded regions indicate the $1\sigma$ credible interval.}
    \label{fig:evidence_comparison}
\end{figure}

\section{Application to Gravitational Wave inference}\label{sec:GW_inf}
In this section we test the performances of \sharpy in handling GW inference problems, while an application to a toy problem is presented in Appendix \ref{sec:11D}. 
For the SMC we fix the number of particles $N_P$ to 5000 and an adaptive temperature scheme with $\alpha = 0.95$ (see section \ref{sec:SMC}). For the NUTS we use a step size $\epsilon = 0.5$ and we let evolve the Hamiltonian system until the ``U-Turn" criterion is reached (see section \ref{sec:NUTS}). We used these settings for all the runs performed in this section, as well as the runs in Appendix \ref{sec:11D}.

We first analyze 100 simulated BBH signals injected in Gaussian noise to verify \sharpy's statistic unbiasedness with the probability-probability test. Then, we test \sharpy on real data, specifically on GW150914,

making a systematic comparison with the results obtained with other reference samplers: two nested sampling algorithms, \texttt{Dynesty}\cite{2020:Dynesty}(via Bilby\cite{2019:bilby_paper}) and \texttt{raynest}\cite{raynest},  and pocoMC\cite{2022:Karamanis_Preconditioned}, a SMC-based algorithm.
\footnote{The settings used for running \texttt{Dynesty}, \texttt{raynest} and \texttt{pocoMC} are available in appendix \ref{sec:settings}.}

Throughout these sections we adopted the aligned-spin waveform model IMRPhenomD \cite{Phenom4, Phenom5} from the \texttt{ripple} package, leading to an 11-dimensional parameter space.
Finally, we run \sharpy on GW150914 with a precessing waveform, IMRPhenomPv2\cite{PhenomPv2}, that increases the dimensionality of the parameter space to 15. 


\subsection*{Simulated BBH signals}\label{sec:simulation}

We inject 100 BBH signals into 4 s of Gaussian noise. The detector network is composed by three interferometers, namely the two LIGO ones and Virgo.
\footnote{The PSD used for this simulation is available at this \href{https://git.ligo.org/lscsoft/lalsuite/-/blob/master/lalsimulation/lib/LIGO-P1200087-v18-aLIGO_DESIGN.txt}{link}.}. The parameters of the simulated BBHs are randomly drawn from the prior, uniform in all source parameters, with the exception of luminosity distance which is log-uniform, resulting in an overall optimal signal-to-noise ratio of $39.7_{-23.1}^{+28.9}$. 
A corner plot of the resulting posterior for one of the events can be found in appendix \ref{sec:corner_appendix}.
In Fig.\,\ref{fig:PPPLOT}, we report the probability-probability test of the statistical robustness of the sampler. For each parameter, we check the percentage of events ($y$-axis) for which the injected value is enclosed in a certain confidence interval ($x$-axis). If the sampler is unbiased, we expect the points to lie along the diagonal, which is indeed the case.

With the settings used for this test, \sharpy produced on average around $15000$ 
posterior samples in slightly  more than 15 minutes 
on a single NVIDIA A100 GPU. 
\begin{figure}[]
    \centering
    \resizebox{0.5\textwidth}{!}{%
  \includegraphics{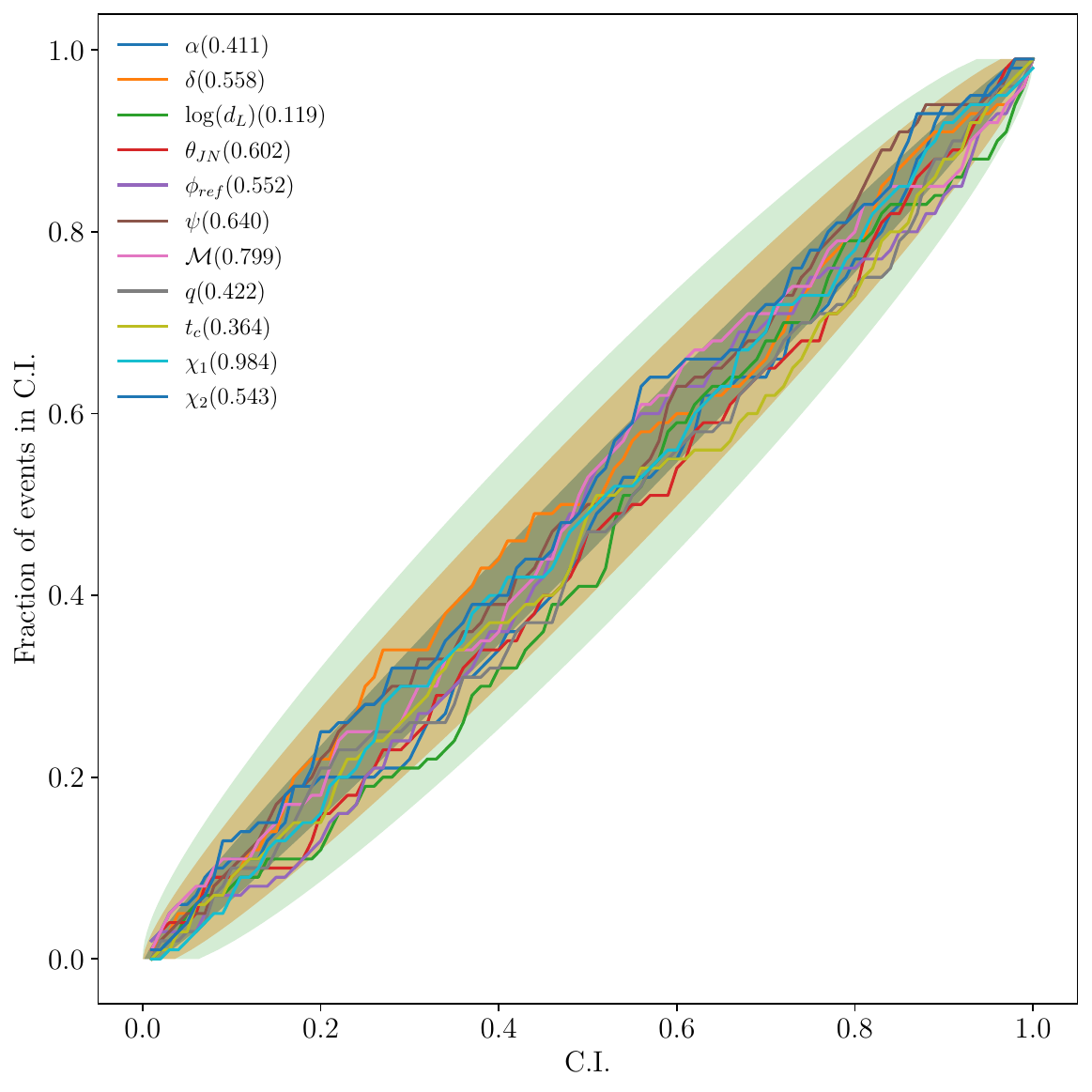}
  }
    \caption{Probability-probability test for the simulated BBH systems. For each parameter of the binary, the plot reports on the $y$-axis the fraction of events for which the true value lies within the credible interval (C.I.) on the $x$-axis. 
    The resulting p-values for each parameters are reported in the corresponding legend entry. The shaded bands represents the 1-2-3 $\sigma$ quantiles. }
    \label{fig:PPPLOT}
\end{figure}

\subsection*{Real data: GW150914}\label{sec:GW150914}
We test  the performance of \sharpy on real data choosing GW150914\cite{150914} as a benchmark and performing 100 independent runs. 
We considered 2 s of data sampled at 1024 Hz.
With the same settings and prior bounds as before, around 17000 samples were produced in about 10 minutes. The number of SMC iterations needed to go from $\beta = 0$ to $\beta = 1$ is around 55. 


Figure \ref{fig:samples_comparison}  shows a comparison between the posteriors obtained with one of the \sharpy runs and the reference samplers, both for the intrinsic parameters (left) and for some of the extrinsic parameters (right).

To quantify the closeness of the two sets of posterior samples, we follow previous literature computing the Jensen-Shannon (JS) divergence, that measures the distance between two probability distributions. It ranges between 0, when the two distributions are equal, and 1, indicating maximum divergence. We compute the JS divergence between the marginal posteriors obtained with  the reference samplers and each of the \sharpy runs. The results are shown in Figure \ref{fig:js_divergence}. For each parameter, the triangles indicate the median value of the JS divergence while the error bars represent the 90\% credible interval.

The JS divergence that we obtain is generally below (or very close to) the threshold proposed in \cite{2020:bilby_ROmero} of 1.5 mb for two sets of samples to be drawn from the same distribution.
However,  for the extrinsic parameters, it tends to be higher compared to other parameters, in particular in the \texttt{Dynesty} case, confirming the general difficulty in sampling the posterior when in the sky parametrization chosen in this work. \cite{2020:bilby_ROmero}.



 Additionally, we test the performance of \sharpy in the computation of the evidence, 
comparing it with the value obtained with the other samplers.

 In Fig.\,\ref{fig:evidence_comparison} we show the distribution of the evidences obtained with 100 independent runs on GW150914 (solid line)
compared with the results obtained with \texttt{Dynesty}, \texttt{raynest} and \texttt{pocoMC}.
The \sharpy evidence distribution is agreement with the other samplers at the 90\% level, confirming the capability of \sharpy of an unbiased computation of the evidence.

\subsection*{GW150914 with a precessing waveform}
For testing the scalability of \sharpy with the number of dimensions, we analyzed GW150914 with IMRPhenomPv2, using the \texttt{ripple} implementation. 
Table \ref{tab:sharpy_gwosc_comparison} shows a comparison, in terms of summary statistics, between the results obtained with \sharpy and GWTC-1, showing a good agreement. The total sampling time is about 20 minutes. 
The full corner plot of the posterior samples obtained with \sharpy can be found in appendix \ref{sec:corner_appendix}.
\renewcommand{\arraystretch}{1.3}
\setlength{\tabcolsep}{8pt}
\begin{table}[ht]

\centering
\begin{tabular}{lcc}
\toprule
Parameter & \sharpy & GWTC-1\\
\midrule
$\mathcal{M} \  [M_\odot]$ & $31.06^{+1.50}_{-1.53}$ & $31.33^{+1.05}_{-1.18}$ \\
$q$ & $0.86^{+0.12}_{-0.20}$ & $0.89^{+0.10}_{-0.19}$ \\
$\chi_{\text{eff}}$ & $-0.03^{+0.11}_{-0.12}$ & $-0.00^{+0.07}_{-0.09}$ \\
$\chi_{p}$ & $0.51^{+0.51}_{-0.39}$ & $0.27^{+0.58}_{-0.20}$ \\
$d_L \ [\mathrm{Mpc}]$ & $445.37^{+143.61}_{-171.09}$ & $437.81^{+145.80}_{-196.65}$ \\
$\theta_{JN}$ & $2.72^{+0.32}_{-0.83}$ & $2.72^{+0.39}_{-1.07}$ \\
$\alpha$ & $1.69^{+0.86}_{-0.74}$ & $1.70^{+0.80}_{-0.73}$ \\
$\delta$ & $-1.21^{+0.23}_{-0.07}$ & $-1.22^{+0.20}_{-0.06}$ \\
\bottomrule
\end{tabular}
\caption{Comparison between \sharpy and GWTC-1 results for GW150914 parameters. Values are reported as median with 90\% credible intervals.}
\label{tab:sharpy_gwosc_comparison}
\end{table}


\section{Conclusions }\label{sec:Conclusions}
In this work we presented \sharpy, a new sampler for gravitational-wave Bayesian inference. It uses the efficient No-U-Turn-Sampler as a mutation kernel of a Sequential Monte Carlo, with the exploration of the parameter space further enhanced by adapting the mass matrix to the local geometry of the distribution. 
Moreover, the \texttt{JAX} implementation allows for a fast and accurate computation of gradients as well as for GPU acceleration, exploiting the intrinsic parallelism of SMC methods.

Remarkably, to the best of our knowledge, this is the first application of the NUTS to single event GW inference problems, carrying with it the efficient parameter space exploration in large-dimensional problems.
We tested the algorithm directly on real data, specifically on GW150914 with an aligned-spin waveform model, demonstrating that \sharpy is able to produce results consistent with other samplers for both the inferred posterior distribution and the evidence value.
Additionally, we performed the probability-probability test, demonstrating the statistical unbiasedness of our sampler.
On a single NVIDIA A100 GPU, the total sampling time is of the order of 10 minutes. 

While is this work we focused on relatively short signals, analyzing signals with arbitrary large durations or frequency resolutions (e.g. signals from binary neutron star mergers) should not in principle impact the performance of \sharpy, since the evaluation of the waveforms happens in parallel. The limited amount of GPU memory however prevents us from using arbitrary large durations or frequency arrays on a single GPU. This can easily mitigated by using frequency bin reduction schemes, that can lower the frequency resolution (and thus the memory required) by order of magnitudes \cite{2020:ROQ, 2021:Multibanding_MOrisaki, 2023:Relative_Binning}. Alternatively, the SMC parallelism can be exploited to split computation and memory across multiples GPUs or the computation can be split in serial batches.

In this work we test \sharpy mainly on an 11-dimensional parameter space, that is typically smaller than standard problems. However,  we do not expect our findings to change significantly in full scale scenarios, since both the SMC and the NUTS, in particular, scale naturally better than standard algorithms with the number of dimensions. Indeed, we partially verified this by analyzing GW150914 with a fully precessing waveform, that increased the dimensionality of the parameter space to 15, obtaining results consistent with the publicly available ones, at a negligible additional cost. A systematic application of \sharpy to precessing systems and a comparison with other samplers covering a broader parameter space is left for future work.
\sharpy is potentially suited for large-dimension problems, such as hierarchical inference, widely used in population and cosmology analyses \cite{2025:full_hier_mancarella, 2025:O4_population, 2025:O4_cosmo} and tests of General Relativity \cite{2021:O3_TGR}, where additional parameters are added to general-relativistic waveforms in order to capture potential deviations and the presence of multiple signals in the data. 
Moreover, the SMC scheme is not limited to problems in which the data and the models are fixed. It can be applied to scenarios where the amount of data to consider varies over time, without the need of repeating the analysis from scratch when new data arrives, e.g. in early-warning and low-latency analysis, where the rapid availability of results is fundamental. Additionally, SMCs can be used to perform inference with a new model starting from the results obtained with a different one \cite{2025:ASPIRE}.
While in this work we used a relatively basic version of the SMC, the efficiency can be improved further by using alternative SMC schemes such as the Persistent Sampling\cite{2024_Persistent}, in which particles are recycled across each iteration potentially leading to an overall performance improvement at no extra computational cost. Further, this scheme should also lead to a reduction in variance both on the inferred posterior distribution and evidence estimate.\\
To conclude, in this work we integrated the No-U-Turn-Sampler into a Sequential Monte Carlo framework, taking advantage of the GPU acceleration and autodifferentiation capabilities of \texttt{JAX}. 
We showed that this combination provides a viable and fast alternative to Nested Sampling, offering the appealing prospect of reducing the computational burden of gravitational-wave inference expected in the near and far future.

\section*{Acknowledgments}
\noindent
We thank the anonymous referee for comments that improved the manuscript.
We thank Michael Williams for providing useful comments on the manuscript and for helpful discussions.\\
We acknowledge ISCRA for awarding this project access to the LEONARDO supercomputer, owned by the EuroHPC Joint Undertaking, hosted by CINECA (Italy).
This work has been supported by the project
BIGA “Boosting Inference for Gravitational-wave Astrophysics” funded by the MUR Progetti di Ricerca di
Rilevante Interesse Nazionale (PRIN) Bando 2022, grant
20228TLHPE, CUP I53D23000630006.\\ GD acknowledges financial support from the National Recovery and
Resilience Plan (PNRR), Mission 4 Component 2 Investment 1.4, National Center for HPC, Big Data and
Quantum Computing, funded by the European Union, NextGenerationEU, CUP B83C22002830001.  FP acknowledges support from the ICSC,  Centro Nazionale di Ricerca in High Performance Computing, Big Data and Quantum Computing, funded by the European Union, NextGenerationEU.
This research has made use of data or software obtained from the Gravitational Wave Open Science Center (gwosc.org), a service of the LIGO Scientific Collaboration, the Virgo Collaboration, and KAGRA. This material is based upon work supported by NSF's LIGO Laboratory which is a major facility fully funded by the National Science Foundation, as well as the Science and Technology Facilities Council (STFC) of the United Kingdom, the Max-Planck-Society (MPS), and the State of Niedersachsen/Germany for support of the construction of Advanced LIGO and construction and operation of the GEO600 detector. Additional support for Advanced LIGO was provided by the Australian Research Council. Virgo is funded, through the European Gravitational Observatory (EGO), by the French Centre National de Recherche Scientifique (CNRS), the Italian Istituto Nazionale di Fisica Nucleare (INFN) and the Dutch Nikhef, with contributions by institutions from Belgium, Germany, Greece, Hungary, Ireland, Japan, Monaco, Poland, Portugal, Spain. KAGRA is supported by Ministry of Education, Culture, Sports, Science and Technology (MEXT), Japan Society for the Promotion of Science (JSPS) in Japan; National Research Foundation (NRF) and Ministry of Science and ICT (MSIT) in Korea; Academia Sinica (AS) and National Science and Technology Council (NSTC) in Taiwan.
\subsubsection*{Software}
This work made use of  
\texttt{JAX}\cite{jax2018github},  \texttt{Bilby}\cite{colm_talbot_2024_14025488}, \texttt{corner}\cite{Foreman-Mackey2016},  
 \texttt{matplotlib}\cite{matplotib},
 \texttt{scipy}\cite{scipy},
 \texttt{numpy}\cite{numpy} and
\texttt{PESummary}\cite{pesummary}.

\appendix
\section{Bimodal 11-D distribution}\label{sec:11D}
We test \sharpy on a bimodal 11-D distribution  $ p(\boldsymbol{x})$ defined as:
\begin{equation}
\label{eq:11Dbimodal}
    p(\boldsymbol{x}) = \mathcal{N}(\boldsymbol{x}|\boldsymbol{\mu}_1, \boldsymbol{\sigma}_1) + \mathcal{N}(\boldsymbol{x}|\boldsymbol{\mu}_2, \boldsymbol{\sigma}_2)
\end{equation}
where $\mathcal{N}$ is a multivariate Gaussian distribution. 
We choose $\boldsymbol{\mu}_1 = \mathbf{1}_{11}$ while $\boldsymbol{\mu}_2 = -\mathbf{1}_{11}$, with $\mathbf{1}_{11}$ indicating an 11-dimensional vector where all the entries are 1. The covariance matrices $\boldsymbol{\sigma_1}, \boldsymbol{\sigma}_2$ are both set to $0.01\mathcal{I}$, where $\mathcal{I}$ is the identity. This results in a distribution with two very distinct peaks. To obtain samples from this distribution we use the same configuration of \sharpy as in Section \ref{sec:GW_inf}.

\begin{figure}[h]
    \centering
   \resizebox{0.5\textwidth}{!}{\includegraphics{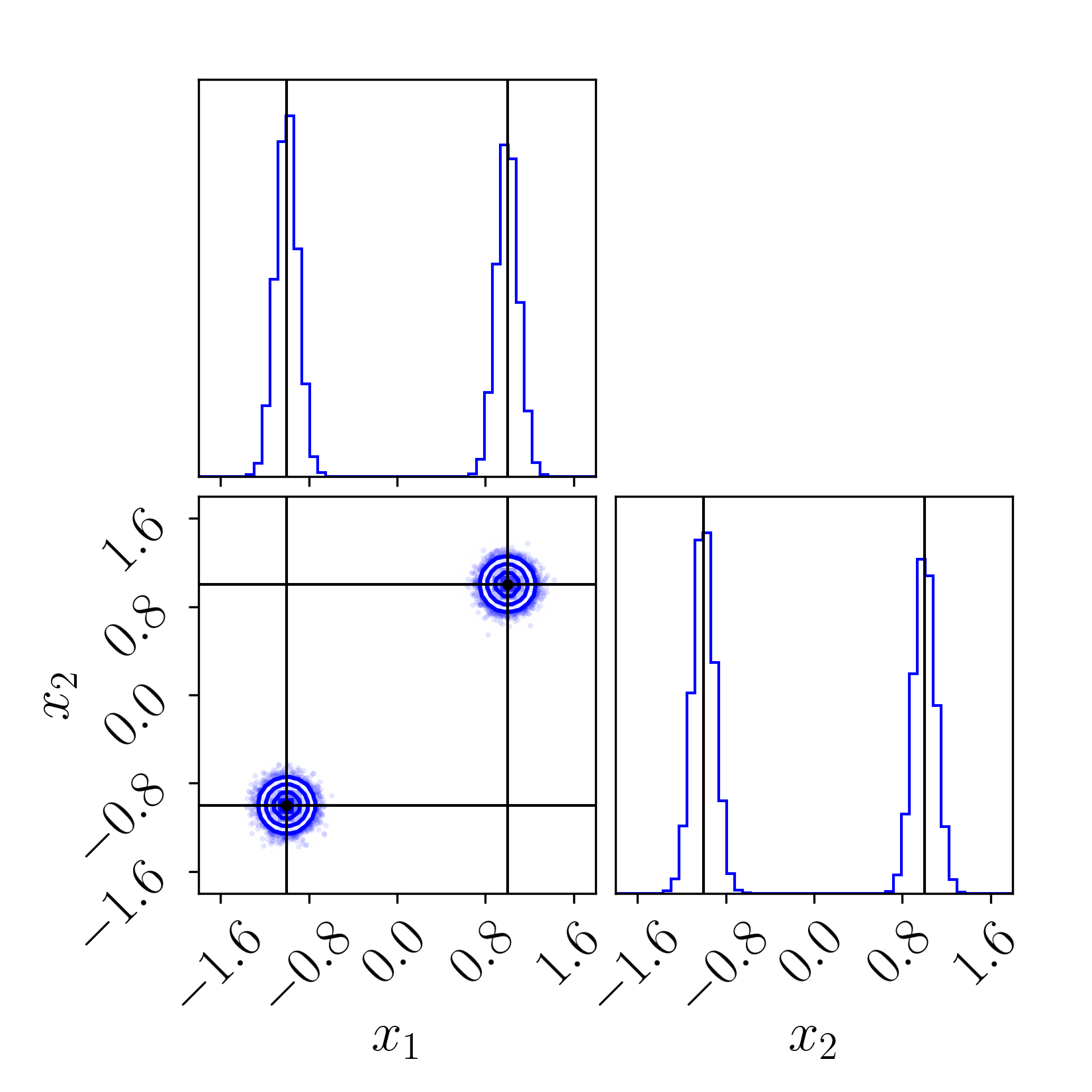}}
    \caption{Marginal samples from the first two dimensions of the bimodal 11-D distribution introduced in Eq.\,\eqref{eq:11Dbimodal}.}
    \label{fig:11D_mixture}
\end{figure}

\begin{figure}[h!]
    \centering
    \resizebox{0.5\textwidth}{!}{\includegraphics{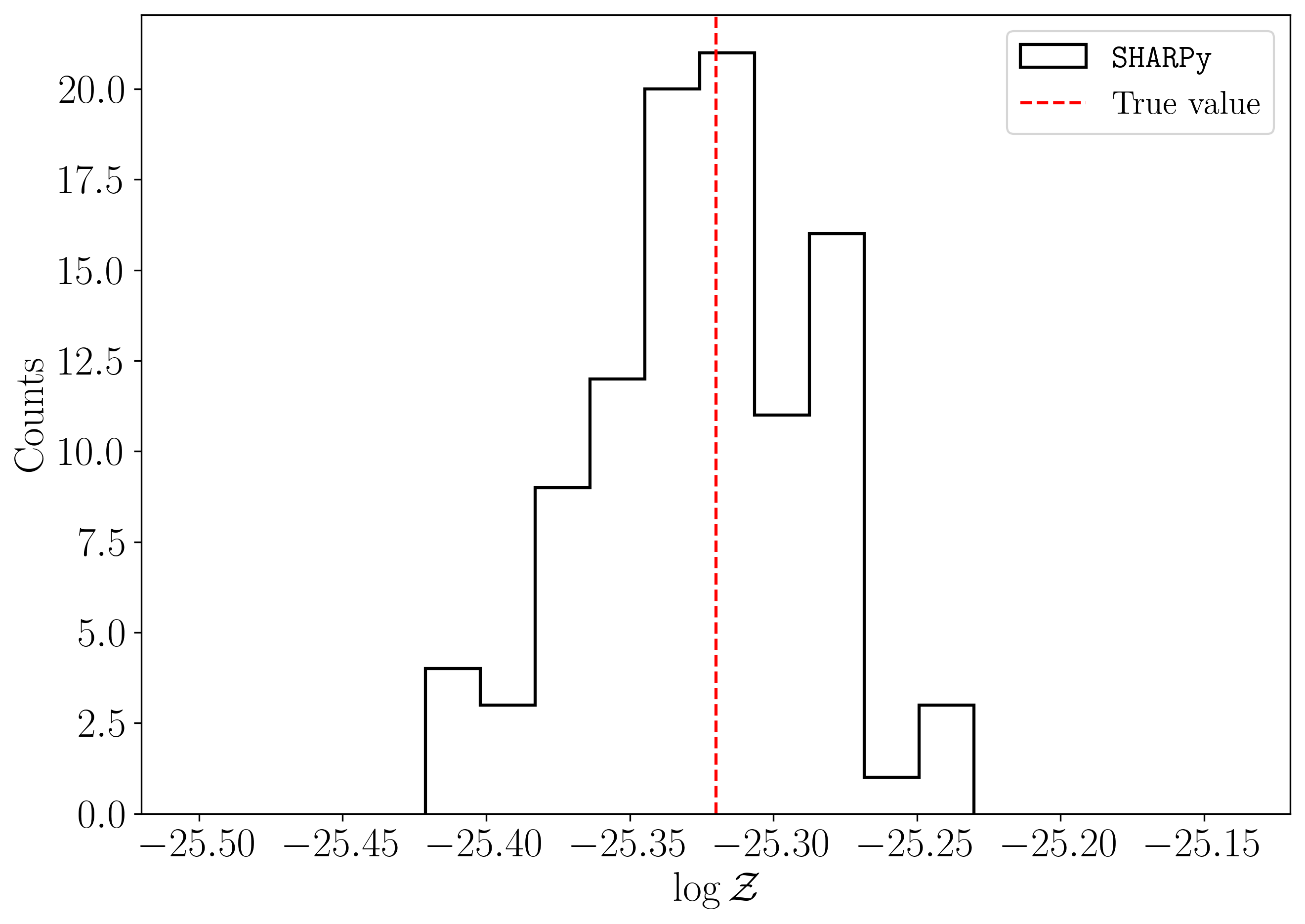}}
    \caption{Histogram of the evidences estimated by 100 independent \sharpy runs compared to the true analytical value.}
    \label{fig:evidence_coparison_mixture}
\end{figure}

In Fig.\,\ref{fig:11D_mixture} we show a corner plot of the posterior samples from the  two-dimensional marginal distribution. 
Additionally, we perform 100 independent \sharpy runs to study the distribution of the evidence, comparing it against the true (and analytic) value. We report the results in Fig.\,\ref{fig:evidence_coparison_mixture}. The distribution obtained is centered around the true value, suggesting no evident biases in the evidence computation, as expected.  

\section{Additional corner plots}\label{sec:corner_appendix}

\begin{figure*}[h]
    \centering
    \resizebox{1\textwidth}{!}{\includegraphics{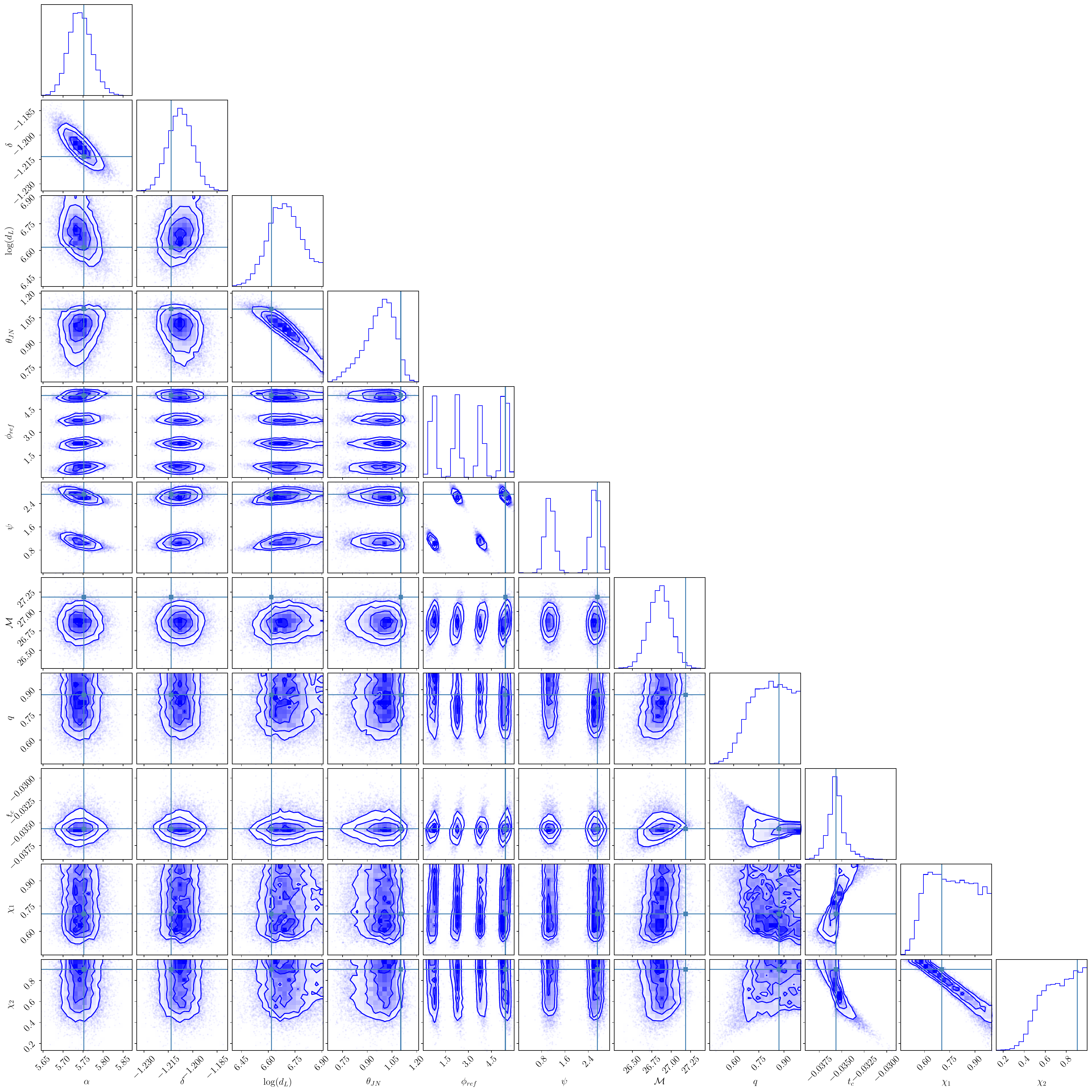}}
    \caption{Corner plot of the samples obtain in one of the injections performed in section \ref{sec:simulation}. The lines indicate the injection parameters.}
    \label{fig:corner_injection}
\end{figure*}

\begin{figure*}[h!]
    \centering
   \resizebox{1\textwidth}{!}{\includegraphics{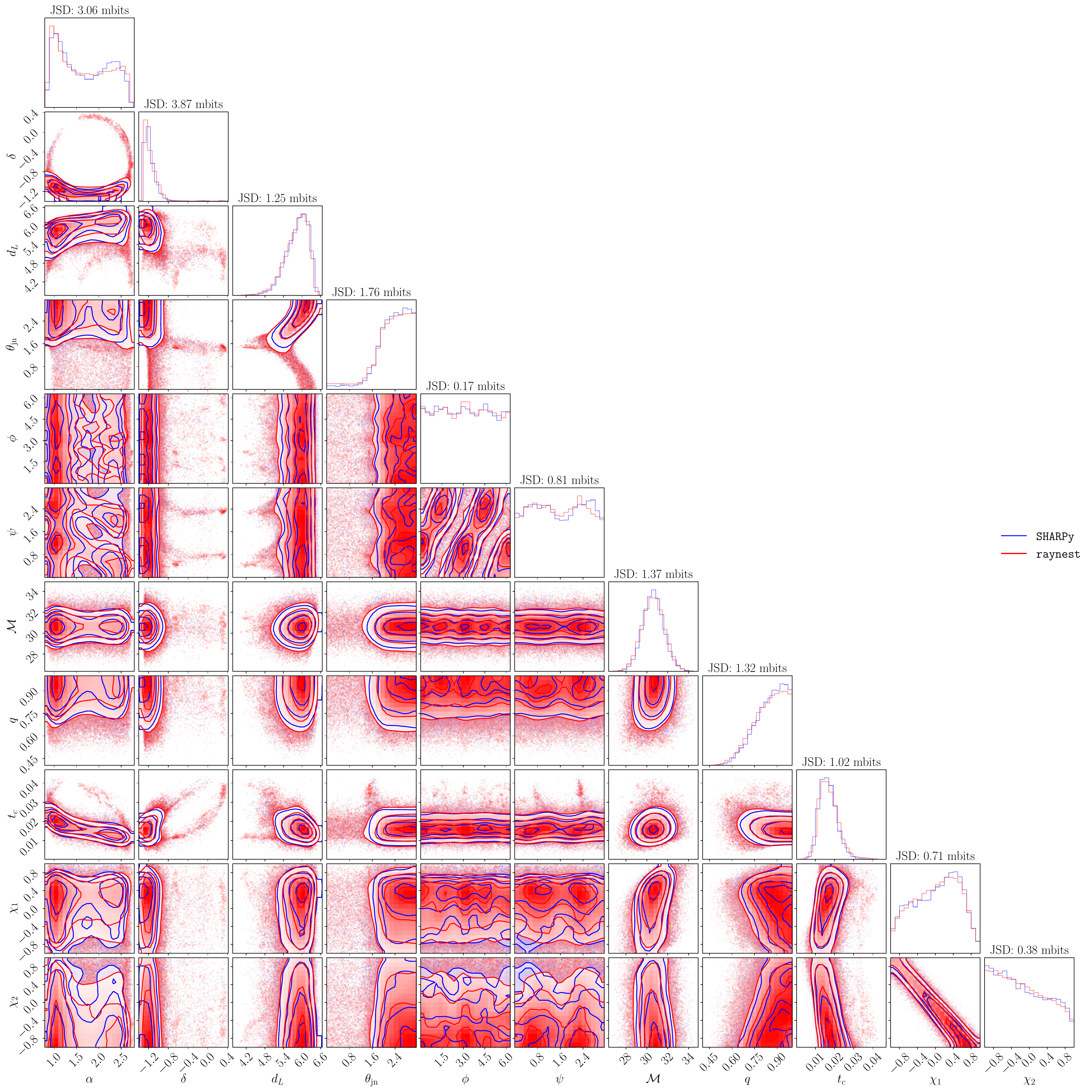}}
    \caption{Full corner plot of the comparison between the samples of GW150914 obtained with \texttt{raynest} and \sharpy, partially showed in fig. \ref{fig:samples_comparison} of section \ref{sec:GW150914}. The value of the JD divergence (JSD) between the two set of samples is reported at the top of each marginal 1D plot in the diagonal. } 
    \label{fig:corner_full_gw}.
\end{figure*}

\begin{figure*}[h!]
    \centering
   \resizebox{1\textwidth}{!}{\includegraphics{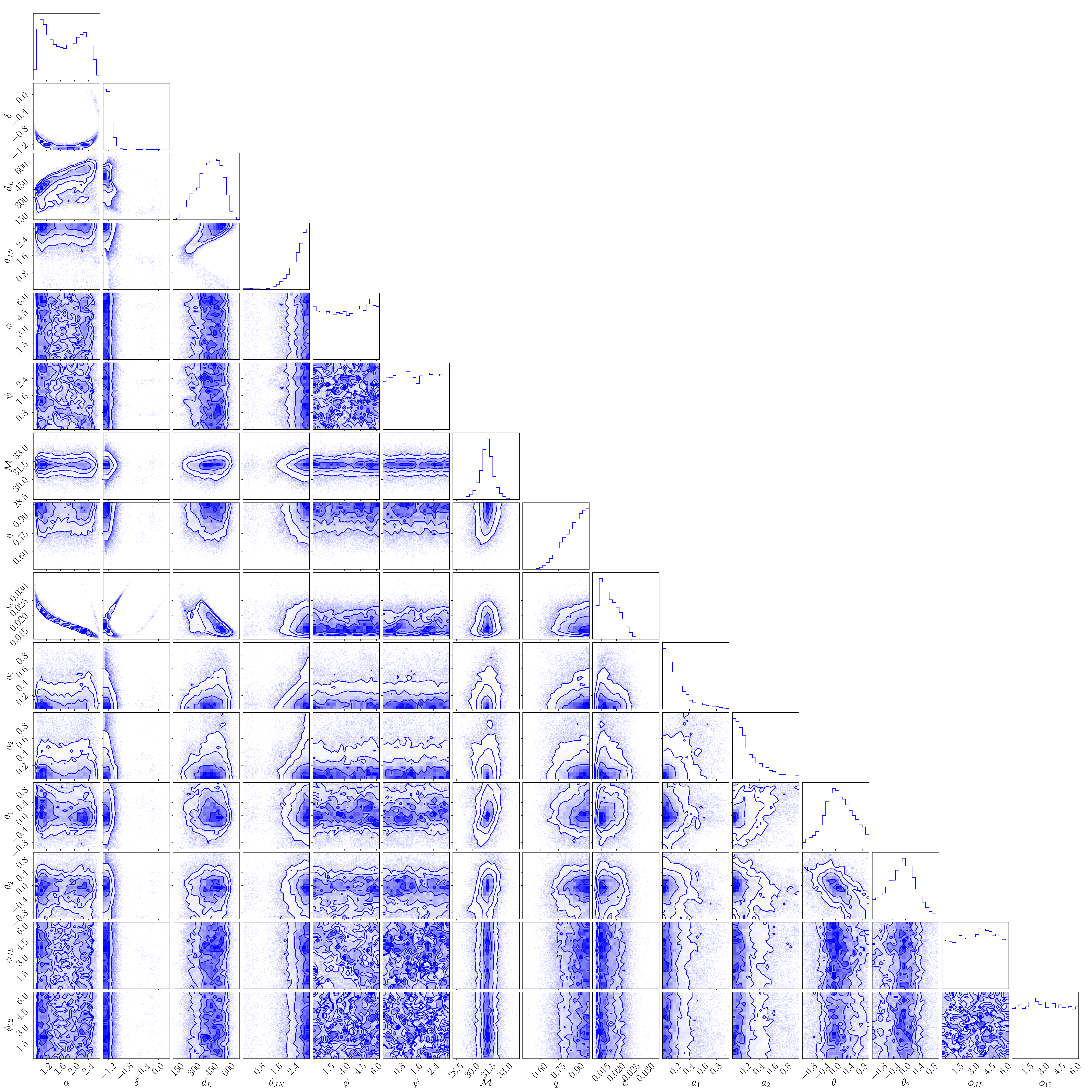}}
    \caption{Corner plot of the posterior samples obtained analyzing GW150914 with the precessing waveform IMRPhenomPv2. } 
    \label{fig:corner_full_gw_PV2}
\end{figure*}

Figure \ref{fig:corner_injection}  
shows the resulting corner plot for one of the injections described in section \ref{sec:simulation}.
Figure \ref{fig:corner_full_gw} shows comparison corner plot between \texttt{Dynesty} and \sharpy including also the three parameters omitted in fig. \ref{fig:samples_comparison}, namely the phase, the polarization angle and the coalescence time. At the top of each marginal 1D plot we show also the JS divergence (JSD) between the two set of samples.
Figure \ref{fig:corner_full_gw_PV2} show the corner plot of the posterior samples of GW150914 obtained with \sharpy and the precessing waveform IMRPhenomPv2.

\section{Reference sampler settings}\label{sec:settings}

In this section we report the settings used for runs of GW150914 with \texttt{Dynesty}, \texttt{raynest} and \texttt{pocoMC} presented in section \ref{sec:GW150914}.
\newline
\texttt{Dynesty}:
\newline
\begin{align*}
    \texttt{nlive} &= 4000;\\
    \texttt{sample} &= \mathrm{acceptance-walk};\\
    \texttt{naccept} &= 60;\\
\texttt{maxmcmc} &= 10000.\\
\end{align*}

\texttt{raynest}:
\newline
\begin{align*}
    \texttt{nnest} &= 6;\\
    \texttt{nensemble} &= 24;\\
    \texttt{nlive} &= 4000;\\
    \texttt{maxmcmc} &= 5000.\\
\end{align*}

\texttt{pocoMC}:
\newline
\begin{align*}
    \texttt{precondition} &= \mathrm{True};\\
    \texttt{n\_effective} &= 2000;\\
    \texttt{n\_active} &= 1000;\\
    \texttt{n\_total} &= 10000;\\
    \texttt{n\_steps} &= 100;\\
    \texttt{n\_max\_steps} &= 1000.\\
\end{align*}

%


\begingroup
\sloppy
\printbibliography[title={References}]

@article{NUTS-2014,
  title={The No-U-Turn sampler: adaptively setting path lengths in Hamiltonian Monte Carlo.},
  author={Hoffman, Matthew D and Gelman, Andrew and others},
  journal={J. Mach. Learn. Res.},
  volume={15},
  number={1},
  pages={1593--1623},
  year={2014}
}

@ARTICLE{2015:LIGO,
       author = {{LIGO Scientific Collaboration} and {Aasi}, J. and {Abbott}, B.~P. and {Abbott}, R. and {Abbott}, T. and {Abernathy}, M.~R. and {Ackley}, K. and {Adams}, C. and {Adams}, T. and {Addesso}, P. and {Adhikari}, R.~X. and {Adya}, V. and {Affeldt}, C. and {Aggarwal}, N. and {Aguiar}, O.~D. and {Ain}, A. and {Ajith}, P. and {Alemic}, A. and {Allen}, B. and {Amariutei}, D. and {Anderson}, S.~B. and {Anderson}, W.~G. and {Arai}, K. and {Araya}, M.~C. and {Arceneaux}, C. and {Areeda}, J.~S. and {Ashton}, G. and {Ast}, S. and {Aston}, S.~M. and {Aufmuth}, P. and {Aulbert}, C. and {Aylott}, B.~E. and {Babak}, S. and {Baker}, P.~T. and {Ballmer}, S.~W. and {Barayoga}, J.~C. and {Barbet}, M. and {Barclay}, S. and {Barish}, B.~C. and {Barker}, D. and {Barr}, B. and {Barsotti}, L. and {Bartlett}, J. and {Barton}, M.~A. and {Bartos}, I. and {Bassiri}, R. and {Batch}, J.~C. and {Baune}, C. and {Behnke}, B. and {Bell}, A.~S. and {Bell}, C. and {Benacquista}, M. and {Bergman}, J. and {Bergmann}, G. and {Berry}, C.~P.~L. and {Betzwieser}, J. and {Bhagwat}, S. and {Bhandare}, R. and {Bilenko}, I.~A. and {Billingsley}, G. and {Birch}, J. and {Biscans}, S. and {Biwer}, C. and {Blackburn}, J.~K. and {Blackburn}, L. and {Blair}, C.~D. and {Blair}, D. and {Bock}, O. and {Bodiya}, T.~P. and {Bojtos}, P. and {Bond}, C. and {Bork}, R. and {Born}, M. and {Bose}, Sukanta and {Brady}, P.~R. and {Braginsky}, V.~B. and {Brau}, J.~E. and {Bridges}, D.~O. and {Brinkmann}, M. and {Brooks}, A.~F. and {Brown}, D.~A. and {Brown}, D.~D. and {Brown}, N.~M. and {Buchman}, S. and {Buikema}, A. and {Buonanno}, A. and {Cadonati}, L. and {Calder{\'o}n Bustillo}, J. and {Camp}, J.~B. and {Cannon}, K.~C. and {Cao}, J. and {Capano}, C.~D. and {Caride}, S. and {Caudill}, S. and {Cavagli{\`a}}, M. and {Cepeda}, C. and {Chakraborty}, R. and {Chalermsongsak}, T. and {Chamberlin}, S.~J. and {Chao}, S. and {Charlton}, P. and {Chen}, Y. and {Cho}, H.~S. and {Cho}, M. and {Chow}, J.~H. and {Christensen}, N. and {Chu}, Q. and {Chung}, S. and {Ciani}, G. and {Clara}, F. and {Clark}, J.~A. and {Collette}, C. and {Cominsky}, L. and {Constancio}, Jr., M. and {Cook}, D. and {Corbitt}, T.~R. and {Cornish}, N. and {Corsi}, A. and {Costa}, C.~A. and {Coughlin}, M.~W. and {Countryman}, S. and {Couvares}, P. and {Coward}, D.~M. and {Cowart}, M.~J. and {Coyne}, D.~C. and {Coyne}, R. and {Craig}, K. and {Creighton}, J.~D.~E. and {Creighton}, T.~D. and {Cripe}, J. and {Crowder}, S.~G. and {Cumming}, A. and {Cunningham}, L. and {Cutler}, C. and {Dahl}, K. and {Dal Canton}, T. and {Damjanic}, M. and {Danilishin}, S.~L. and {Danzmann}, K. and {Dartez}, L. and {Dave}, I. and {Daveloza}, H. and {Davies}, G.~S. and {Daw}, E.~J. and {DeBra}, D. and {Del Pozzo}, W. and {Denker}, T. and {Dent}, T. and {Dergachev}, V. and {DeRosa}, R.~T. and {DeSalvo}, R. and {Dhurandhar}, S. and {D{\textasciiacute}{\i}az}, M. and {Di Palma}, I. and {Dojcinoski}, G. and {Dominguez}, E. and {Donovan}, F. and {Dooley}, K.~L. and {Doravari}, S. and {Douglas}, R. and {Downes}, T.~P. and {Driggers}, J.~C. and {Du}, Z. and {Dwyer}, S. and {Eberle}, T. and {Edo}, T. and {Edwards}, M. and {Edwards}, M. and {Effler}, A. and {Eggenstein}, H.-B. and {Ehrens}, P. and {Eichholz}, J. and {Eikenberry}, S.~S. and {Essick}, R. and {Etzel}, T. and {Evans}, M. and {Evans}, T. and {Factourovich}, M. and {Fairhurst}, S. and {Fan}, X. and {Fang}, Q. and {Farr}, B. and {Farr}, W.~M. and {Favata}, M. and {Fays}, M. and {Fehrmann}, H. and {Fejer}, M.~M. and {Feldbaum}, D. and {Ferreira}, E.~C. and {Fisher}, R.~P. and {Frei}, Z. and {Freise}, A. and {Frey}, R. and {Fricke}, T.~T. and {Fritschel}, P. and {Frolov}, V.~V. and {Fuentes-Tapia}, S. and {Fulda}, P. and {Fyffe}, M. and {Gair}, J.~R.},
        title = "{Advanced LIGO}",
      journal = {Classical and Quantum Gravity},
         year = 2015,
        month = apr,
       volume = {32},
       number = {7},
          eid = {074001},
        pages = {074001},
          doi = {10.1088/0264-9381/32/7/074001},
archivePrefix = {arXiv},
       eprint = {1411.4547},
 primaryClass = {gr-qc},
       adsurl = {https://ui.adsabs.harvard.edu/abs/2015CQGra..32g4001L}
}

@ARTICLE{2015:VIRGO,
       author = {{Acernese}, F. and {Agathos}, M. and {Agatsuma}, K. and {Aisa}, D. and {Allemandou}, N. and {Allocca}, A. and {Amarni}, J. and {Astone}, P. and {Balestri}, G. and {Ballardin}, G. and {Barone}, F. and {Baronick}, J.-P. and {Barsuglia}, M. and {Basti}, A. and {Basti}, F. and {Bauer}, Th S. and {Bavigadda}, V. and {Bejger}, M. and {Beker}, M.~G. and {Belczynski}, C. and {Bersanetti}, D. and {Bertolini}, A. and {Bitossi}, M. and {Bizouard}, M.~A. and {Bloemen}, S. and {Blom}, M. and {Boer}, M. and {Bogaert}, G. and {Bondi}, D. and {Bondu}, F. and {Bonelli}, L. and {Bonnand}, R. and {Boschi}, V. and {Bosi}, L. and {Bouedo}, T. and {Bradaschia}, C. and {Branchesi}, M. and {Briant}, T. and {Brillet}, A. and {Brisson}, V. and {Bulik}, T. and {Bulten}, H.~J. and {Buskulic}, D. and {Buy}, C. and {Cagnoli}, G. and {Calloni}, E. and {Campeggi}, C. and {Canuel}, B. and {Carbognani}, F. and {Cavalier}, F. and {Cavalieri}, R. and {Cella}, G. and {Cesarini}, E. and {Mottin}, E. Chassande- and {Chincarini}, A. and {Chiummo}, A. and {Chua}, S. and {Cleva}, F. and {Coccia}, E. and {Cohadon}, P.-F. and {Colla}, A. and {Colombini}, M. and {Conte}, A. and {Coulon}, J.-P. and {Cuoco}, E. and {Dalmaz}, A. and {D'Antonio}, S. and {Dattilo}, V. and {Davier}, M. and {Day}, R. and {Debreczeni}, G. and {Degallaix}, J. and {Del{\'e}glise}, S. and {Pozzo}, W. Del and {Dereli}, H. and {Rosa}, R. De and {Fiore}, L. Di and {Lieto}, A. Di and {Virgilio}, A. Di and {Doets}, M. and {Dolique}, V. and {Drago}, M. and {Ducrot}, M. and {Endr{\H{o}}czi}, G. and {Fafone}, V. and {Farinon}, S. and {Ferrante}, I. and {Ferrini}, F. and {Fidecaro}, F. and {Fiori}, I. and {Flaminio}, R. and {Fournier}, J.-D. and {Franco}, S. and {Frasca}, S. and {Frasconi}, F. and {Gammaitoni}, L. and {Garufi}, F. and {Gaspard}, M. and {Gatto}, A. and {Gemme}, G. and {Gendre}, B. and {Genin}, E. and {Gennai}, A. and {Ghosh}, S. and {Giacobone}, L. and {Giazotto}, A. and {Gouaty}, R. and {Granata}, M. and {Greco}, G. and {Groot}, P. and {Guidi}, G.~M. and {Harms}, J. and {Heidmann}, A. and {Heitmann}, H. and {Hello}, P. and {Hemming}, G. and {Hennes}, E. and {Hofman}, D. and {Jaranowski}, P. and {Jonker}, R.~J.~G. and {Kasprzack}, M. and {K{\'e}f{\'e}lian}, F. and {Kowalska}, I. and {Kraan}, M. and {Kr{\'o}lak}, A. and {Kutynia}, A. and {Lazzaro}, C. and {Leonardi}, M. and {Leroy}, N. and {Letendre}, N. and {Li}, T.~G.~F. and {Lieunard}, B. and {Lorenzini}, M. and {Loriette}, V. and {Losurdo}, G. and {Magazz{\`u}}, C. and {Majorana}, E. and {Maksimovic}, I. and {Malvezzi}, V. and {Man}, N. and {Mangano}, V. and {Mantovani}, M. and {Marchesoni}, F. and {Marion}, F. and {Marque}, J. and {Martelli}, F. and {Martellini}, L. and {Masserot}, A. and {Meacher}, D. and {Meidam}, J. and {Mezzani}, F. and {Michel}, C. and {Milano}, L. and {Minenkov}, Y. and {Moggi}, A. and {Mohan}, M. and {Montani}, M. and {Morgado}, N. and {Mours}, B. and {Mul}, F. and {Nagy}, M.~F. and {Nardecchia}, I. and {Naticchioni}, L. and {Nelemans}, G. and {Neri}, I. and {Neri}, M. and {Nocera}, F. and {Pacaud}, E. and {Palomba}, C. and {Paoletti}, F. and {Paoli}, A. and {Pasqualetti}, A. and {Passaquieti}, R. and {Passuello}, D. and {Perciballi}, M. and {Petit}, S. and {Pichot}, M. and {Piergiovanni}, F. and {Pillant}, G. and {Piluso}, A. and {Pinard}, L. and {Poggiani}, R. and {Prijatelj}, M. and {Prodi}, G.~A. and {Punturo}, M. and {Puppo}, P. and {Rabeling}, D.~S. and {R{\'a}cz}, I. and {Rapagnani}, P. and {Razzano}, M. and {Re}, V. and {Regimbau}, T. and {Ricci}, F. and {Robinet}, F. and {Rocchi}, A. and {Rolland}, L. and {Romano}, R. and {Rosi{\'n}ska}, D. and {Ruggi}, P. and {Saracco}, E.},
        title = "{Advanced Virgo: a second-generation interferometric gravitational wave detector}",
      journal = {Classical and Quantum Gravity},
         year = 2015,
        month = jan,
       volume = {32},
       number = {2},
          eid = {024001},
        pages = {024001},
          doi = {10.1088/0264-9381/32/2/024001},
archivePrefix = {arXiv},
       eprint = {1408.3978},
 primaryClass = {gr-qc},
       adsurl = {https://ui.adsabs.harvard.edu/abs/2015CQGra..32b4001A}
}

@ARTICLE{2013:KAGRA,
       author = {{Aso}, Yoichi and {Michimura}, Yuta and {Somiya}, Kentaro and {Ando}, Masaki and {Miyakawa}, Osamu and {Sekiguchi}, Takanori and {Tatsumi}, Daisuke and {Yamamoto}, Hiroaki},
        title = "{Interferometer design of the KAGRA gravitational wave detector}",
      journal = {Phys. Rev. D},
         year = 2013,
        month = aug,
       volume = {88},
       number = {4},
          eid = {043007},
        pages = {043007},
          doi = {10.1103/PhysRevD.88.043007},
archivePrefix = {arXiv},
       eprint = {1306.6747},
 primaryClass = {gr-qc},
       adsurl = {https://ui.adsabs.harvard.edu/abs/2013PhRvD..88d3007A}
}

@ARTICLE{2025:O4_population,
       author = {{The LIGO Scientific Collaboration} and {the Virgo Collaboration} and {the KAGRA Collaboration} and {Abac}, A.~G. and {Abouelfettouh}, I. and {Acernese}, F. and {Ackley}, K. and {Adamcewicz}, C. and {Adhicary}, S. and {Adhikari}, D. and {Adhikari}, N. and {Adhikari}, R.~X. and {Adkins}, V.~K. and {Afroz}, S. and {Agarwal}, D. and {Agathos}, M. and {Aghaei Abchouyeh}, M. and {Aguiar}, O.~D. and {Ahmadzadeh}, S. and {Aiello}, L. and {Ain}, A. and {Ajith}, P. and {Akutsu}, T. and {Albanesi}, S. and {Alfaidi}, R.~A. and {Al-Jodah}, A. and {All{\'e}n{\'e}}, C. and {Allocca}, A. and {Al-Shammari}, S. and {Altin}, P.~A. and {Alvarez-Lopez}, S. and {Amarasinghe}, O. and {Amato}, A. and {Amra}, C. and {Ananyeva}, A. and {Anderson}, S.~B. and {Anderson}, W.~G. and {Andia}, M. and {Ando}, M. and {Andrade}, T. and {Andr{\'e}s-Carcasona}, M. and {Andri{\'c}}, T. and {Anglin}, J. and {Ansoldi}, S. and {Antelis}, J.~M. and {Antier}, S. and {Aoumi}, M. and {Appavuravther}, E.~Z. and {Appert}, S. and {Apple}, S.~K. and {Arai}, K. and {Araya}, A. and {Araya}, M.~C. and {Arca Sedda}, M. and {Areeda}, J.~S. and {Argianas}, L. and {Aritomi}, N. and {Armato}, F. and {Armstrong}, S. and {Arnaud}, N. and {Arogeti}, M. and {Aronson}, S.~M. and {Arun}, K.~G. and {Ashton}, G. and {Aso}, Y. and {Assiduo}, M. and {Assis de Souza Melo}, S. and {Aston}, S.~M. and {Astone}, P. and {Attadio}, F. and {Aubin}, F. and {AultONeal}, K. and {Avallone}, G. and {Babak}, S. and {Badaracco}, F. and {Badger}, C. and {Bae}, S. and {Bagnasco}, S. and {Bagui}, E. and {Baiotti}, L. and {Bajpai}, R. and {Baka}, T. and {Baker}, T. and {Ball}, M. and {Ballardin}, G. and {Ballmer}, S.~W. and {Banagiri}, S. and {Banerjee}, B. and {Bankar}, D. and {Baptiste}, T.~M. and {Baral}, P. and {Barayoga}, J.~C. and {Barish}, B.~C. and {Barker}, D. and {Barman}, N. and {Barneo}, P. and {Barone}, F. and {Barr}, B. and {Barsotti}, L. and {Barsuglia}, M. and {Barta}, D. and {Bartoletti}, A.~M. and {Barton}, M.~A. and {Bartos}, I. and {Basak}, S. and {Basalaev}, A. and {Bassiri}, R. and {Basti}, A. and {Bates}, D.~E. and {Bawaj}, M. and {Baxi}, P. and {Bayley}, J.~C. and {Baylor}, A.~C. and {Baynard}, II, P.~A. and {Bazzan}, M. and {Bedakihale}, V.~M. and {Beirnaert}, F. and {Bejger}, M. and {Belardinelli}, D. and {Bell}, A.~S. and {Bellie}, D.~S. and {Bellizzi}, L. and {Beltran-Martinez}, D. and {Benoit}, W. and {Bentara}, I. and {Bentley}, J.~D. and {Ben Yaala}, M. and {Bera}, S. and {Bergamin}, F. and {Berger}, B.~K. and {Bernuzzi}, S. and {Beroiz}, M. and {Berry}, C.~P.~L. and {Bersanetti}, D. and {Bertolini}, A. and {Betzwieser}, J. and {Beveridge}, D. and {Bevilacqua}, G. and {Bevins}, N. and {Bhandare}, R. and {Bhatt}, R. and {Bhattacharjee}, D. and {Bhaumik}, S. and {Bhowmick}, S. and {Biancalana}, V. and {Bianchi}, A. and {Bilenko}, I.~A. and {Billingsley}, G. and {Binetti}, A. and {Bini}, S. and {Binu}, C. and {Birnholtz}, O. and {Biscoveanu}, S. and {Bisht}, A. and {Bitossi}, M. and {Bizouard}, M. -A. and {Blaber}, S. and {Blackburn}, J.~K. and {Blagg}, L.~A. and {Blair}, C.~D. and {Blair}, D.~G. and {Bobba}, F. and {Bode}, N. and {Boileau}, G. and {Boldrini}, M. and {Bolingbroke}, G.~N. and {Bolliand}, A. and {Bonavena}, L.~D. and {Bondarescu}, R. and {Bondu}, F. and {Bonilla}, E. and {Bonilla}, M.~S. and {Bonino}, A. and {Bonnand}, R. and {Booker}, P. and {Borchers}, A. and {Borhanian}, S. and {Boschi}, V. and {Bose}, S. and {Bossilkov}, V. and {Boudon}, A. and {Bozzi}, A. and {Bradaschia}, C. and {Brady}, P.~R. and {Branch}, A. and {Branchesi}, M. and {Braun}, I. and {Briant}, T. and {Brillet}, A. and {Brinkmann}, M. and {Brockill}, P. and {Brockmueller}, E. and {Brooks}, A.~F. and {Brown}, B.~C. and {Brown}, D.~D. and {Brozzetti}, M.~L. and {Brunett}, S. and {Bruno}, G. and {Bruntz}, R. and {Bryant}, J.},
        title = "{GWTC-4.0: Population Properties of Merging Compact Binaries}",
      journal = {arXiv e-prints},
         year = 2025,
        month = aug,
          eid = {arXiv:2508.18083},
        pages = {arXiv:2508.18083},
          doi = {10.48550/arXiv.2508.18083},
archivePrefix = {arXiv},
       eprint = {2508.18083},
 primaryClass = {astro-ph.HE},
       adsurl = {https://ui.adsabs.harvard.edu/abs/2025arXiv250818083T}
}

@ARTICLE{2025:O4_cosmo,
       author = {{The LIGO Scientific Collaboration} and {the Virgo Collaboration} and {the KAGRA Collaboration} and {Abac}, A.~G. and {Abouelfettouh}, I. and {Acernese}, F. and {Ackley}, K. and {Adamcewicz}, C. and {Adhicary}, S. and {Adhikari}, D. and {Adhikari}, N. and {Adhikari}, R.~X. and {Adkins}, V.~K. and {Afroz}, S. and {Agapito}, A. and {Agarwal}, D. and {Agathos}, M. and {Aggarwal}, N. and {Aggarwal}, S. and {Aguiar}, O.~D. and {Ahrend}, I. -L. and {Aiello}, L. and {Ain}, A. and {Ajith}, P. and {Akutsu}, T. and {Albanesi}, S. and {Ali}, W. and {Al-Kershi}, S. and {All{\'e}n{\'e}}, C. and {Allocca}, A. and {Al-Shammari}, S. and {Altin}, P.~A. and {Alvarez-Lopez}, S. and {Amar}, W. and {Amarasinghe}, O. and {Amato}, A. and {Amicucci}, F. and {Amra}, C. and {Ananyeva}, A. and {Anderson}, S.~B. and {Anderson}, W.~G. and {Andia}, M. and {Ando}, M. and {Andr{\'e}s-Carcasona}, M. and {Andri{\'c}}, T. and {Anglin}, J. and {Ansoldi}, S. and {Antelis}, J.~M. and {Antier}, S. and {Aoumi}, M. and {Appavuravther}, E.~Z. and {Appert}, S. and {Apple}, S.~K. and {Arai}, K. and {Araya}, A. and {Araya}, M.~C. and {Arca Sedda}, M. and {Areeda}, J.~S. and {Aritomi}, N. and {Armato}, F. and {Armstrong}, S. and {Arnaud}, N. and {Arogeti}, M. and {Aronson}, S.~M. and {Arun}, K.~G. and {Ashton}, G. and {Aso}, Y. and {Asprea}, L. and {Assiduo}, M. and {Assis de Souza Melo}, S. and {Aston}, S.~M. and {Astone}, P. and {Attadio}, F. and {Aubin}, F. and {AultONeal}, K. and {Avallone}, G. and {Avila}, E.~A. and {Babak}, S. and {Badger}, C. and {Bae}, S. and {Bagnasco}, S. and {Baiotti}, L. and {Bajpai}, R. and {Baka}, T. and {Baker}, A.~M. and {Baker}, K.~A. and {Baker}, T. and {Baldi}, G. and {Baldicchi}, N. and {Ball}, M. and {Ballardin}, G. and {Ballmer}, S.~W. and {Banagiri}, S. and {Banerjee}, B. and {Bankar}, D. and {Baptiste}, T.~M. and {Baral}, P. and {Baratti}, M. and {Barayoga}, J.~C. and {Barish}, B.~C. and {Barker}, D. and {Barman}, N. and {Barneo}, P. and {Barone}, F. and {Barr}, B. and {Barsotti}, L. and {Barsuglia}, M. and {Barta}, D. and {Bartoletti}, A.~M. and {Barton}, M.~A. and {Bartos}, I. and {Basalaev}, A. and {Bassiri}, R. and {Basti}, A. and {Bawaj}, M. and {Baxi}, P. and {Bayley}, J.~C. and {Baylor}, A.~C. and {Baynard}, II, P.~A. and {Bazzan}, M. and {Bedakihale}, V.~M. and {Beirnaert}, F. and {Bejger}, M. and {Belardinelli}, D. and {Bell}, A.~S. and {Bellie}, D.~S. and {Bellizzi}, L. and {Benoit}, W. and {Bentara}, I. and {Bentley}, J.~D. and {Ben Yaala}, M. and {Bera}, S. and {Bergamin}, F. and {Berger}, B.~K. and {Bernuzzi}, S. and {Beroiz}, M. and {Berry}, C.~P.~L. and {Bersanetti}, D. and {Bertheas}, T. and {Bertolini}, A. and {Betzwieser}, J. and {Beveridge}, D. and {Bevilacqua}, G. and {Bevins}, N. and {Bhandare}, R. and {Bhatt}, R. and {Bhattacharjee}, D. and {Bhattacharyya}, S. and {Bhaumik}, S. and {Biancalana}, V. and {Bianchi}, A. and {Bilenko}, I.~A. and {Bilicki}, M. and {Billingsley}, G. and {Binetti}, A. and {Bini}, S. and {Binu}, C. and {Biot}, S. and {Birnholtz}, O. and {Biscoveanu}, S. and {Bisht}, A. and {Bitossi}, M. and {Bizouard}, M. -A. and {Blaber}, S. and {Blackburn}, J.~K. and {Blagg}, L.~A. and {Blair}, C.~D. and {Blair}, D.~G. and {Bode}, N. and {Boettner}, N. and {Boileau}, G. and {Boldrini}, M. and {Bolingbroke}, G.~N. and {Bolliand}, A. and {Bonavena}, L.~D. and {Bondarescu}, R. and {Bondu}, F. and {Bonilla}, E. and {Bonilla}, M.~S. and {Bonino}, A. and {Bonnand}, R. and {Borchers}, A. and {Borhanian}, S. and {Boschi}, V. and {Bose}, S. and {Bossilkov}, V. and {Bothra}, Y. and {Boudon}, A. and {Bourg}, L. and {Boyle}, M. and {Bozzi}, A. and {Bradaschia}, C. and {Brady}, P.~R. and {Branch}, A. and {Branchesi}, M. and {Braun}, I. and {Briant}, T. and {Brillet}, A. and {Brinkmann}, M. and {Brockill}, P.},
        title = "{GWTC-4.0: Constraints on the Cosmic Expansion Rate and Modified Gravitational-wave Propagation}",
      journal = {arXiv e-prints},
         year = 2025,
        month = sep,
          eid = {arXiv:2509.04348},
        pages = {arXiv:2509.04348},
          doi = {10.48550/arXiv.2509.04348},
archivePrefix = {arXiv},
       eprint = {2509.04348},
 primaryClass = {astro-ph.CO},
       adsurl = {https://ui.adsabs.harvard.edu/abs/2025arXiv250904348T}
}

@ARTICLE{2021:O3_TGR,
       author = {{The LIGO Scientific Collaboration} and {the Virgo Collaboration} and {the KAGRA Collaboration} and {Abbott}, R. and {Abe}, H. and {Acernese}, F. and {Ackley}, K. and {Adhikari}, N. and {Adhikari}, R.~X. and {Adkins}, V.~K. and {Adya}, V.~B. and {Affeldt}, C. and {Agarwal}, D. and {Agathos}, M. and {Agatsuma}, K. and {Aggarwal}, N. and {Aguiar}, O.~D. and {Aiello}, L. and {Ain}, A. and {Ajith}, P. and {Akutsu}, T. and {de Alarc{\'o}n}, P.~F. and {Albanesi}, S. and {Alfaidi}, R.~A. and {Allocca}, A. and {Altin}, P.~A. and {Amato}, A. and {Anand}, C. and {Anand}, S. and {Ananyeva}, A. and {Anderson}, S.~B. and {Anderson}, W.~G. and {Ando}, M. and {Andrade}, T. and {Andres}, N. and {Andr{\'e}s-Carcasona}, M. and {Andri{\'c}}, T. and {Angelova}, S.~V. and {Ansoldi}, S. and {Antelis}, J.~M. and {Antier}, S. and {Apostolatos}, T. and {Appavuravther}, E.~Z. and {Appert}, S. and {Apple}, S.~K. and {Arai}, K. and {Araya}, A. and {Araya}, M.~C. and {Areeda}, J.~S. and {Ar{\`e}ne}, M. and {Aritomi}, N. and {Arnaud}, N. and {Arogeti}, M. and {Aronson}, S.~M. and {Arun}, K.~G. and {Asada}, H. and {Asali}, Y. and {Ashton}, G. and {Aso}, Y. and {Assiduo}, M. and {Assis de Souza Melo}, S. and {Aston}, S.~M. and {Astone}, P. and {Aubin}, F. and {AultONeal}, K. and {Austin}, C. and {Babak}, S. and {Badaracco}, F. and {Bader}, M.~K.~M. and {Badger}, C. and {Bae}, S. and {Bae}, Y. and {Baer}, A.~M. and {Bagnasco}, S. and {Bai}, Y. and {Baird}, J. and {Bajpai}, R. and {Baka}, T. and {Ball}, M. and {Ballardin}, G. and {Ballmer}, S.~W. and {Balsamo}, A. and {Baltus}, G. and {Banagiri}, S. and {Banerjee}, B. and {Bankar}, D. and {Barayoga}, J.~C. and {Barbieri}, C. and {Barish}, B.~C. and {Barker}, D. and {Barneo}, P. and {Barone}, F. and {Barr}, B. and {Barsotti}, L. and {Barsuglia}, M. and {Barta}, D. and {Bartlett}, J. and {Barton}, M.~A. and {Bartos}, I. and {Basak}, S. and {Bassiri}, R. and {Basti}, A. and {Bawaj}, M. and {Bayley}, J.~C. and {Bazzan}, M. and {Becher}, B.~R. and {B{\'e}csy}, B. and {Bedakihale}, V.~M. and {Beirnaert}, F. and {Bejger}, M. and {Belahcene}, I. and {Benedetto}, V. and {Beniwal}, D. and {Benjamin}, M.~G. and {Bennett}, T.~F. and {Bentley}, J.~D. and {BenYaala}, M. and {Bera}, S. and {Berbel}, M. and {Bergamin}, F. and {Berger}, B.~K. and {Bernuzzi}, S. and {Berry}, C.~P.~L. and {Bersanetti}, D. and {Bertolini}, A. and {Betzwieser}, J. and {Beveridge}, D. and {Bhandare}, R. and {Bhandari}, A.~V. and {Bhardwaj}, U. and {Bhatt}, R. and {Bhattacharjee}, D. and {Bhaumik}, S. and {Bianchi}, A. and {Bilenko}, I.~A. and {Billingsley}, G. and {Bini}, S. and {Birney}, R. and {Birnholtz}, O. and {Biscans}, S. and {Bischi}, M. and {Biscoveanu}, S. and {Bisht}, A. and {Biswas}, B. and {Bitossi}, M. and {Bizouard}, M. -A. and {Blackburn}, J.~K. and {Blair}, C.~D. and {Blair}, D.~G. and {Blair}, R.~M. and {Bobba}, F. and {Bode}, N. and {Bo{\"e}r}, M. and {Bogaert}, G. and {Boldrini}, M. and {Bolingbroke}, G.~N. and {Bonavena}, L.~D. and {Bondu}, F. and {Bonilla}, E. and {Bonnand}, R. and {Booker}, P. and {Boom}, B.~A. and {Bork}, R. and {Boschi}, V. and {Bose}, N. and {Bose}, S. and {Bossilkov}, V. and {Boudart}, V. and {Bouffanais}, Y. and {Bozzi}, A. and {Bradaschia}, C. and {Brady}, P.~R. and {Bramley}, A. and {Branch}, A. and {Branchesi}, M. and {Brau}, J.~E. and {Breschi}, M. and {Briant}, T. and {Briggs}, J.~H. and {Brillet}, A. and {Brinkmann}, M. and {Brockill}, P. and {Brooks}, A.~F. and {Brooks}, J. and {Brown}, D.~D. and {Brunett}, S. and {Bruno}, G. and {Bruntz}, R. and {Bryant}, J. and {Bucci}, F. and {Bulik}, T. and {Bulten}, H.~J. and {Buonanno}, A. and {Burtnyk}, K. and {Buscicchio}, R. and {Buskulic}, D. and {Buy}, C. and {Byer}, R.~L. and {Cabourn Davies}, G.~S. and {Cabras}, G.},
        title = "{Tests of General Relativity with GWTC-3}",
      journal = {arXiv e-prints},
         year = 2021,
        month = dec,
          eid = {arXiv:2112.06861},
        pages = {arXiv:2112.06861},
          doi = {10.48550/arXiv.2112.06861},
archivePrefix = {arXiv},
       eprint = {2112.06861},
 primaryClass = {gr-qc},
       adsurl = {https://ui.adsabs.harvard.edu/abs/2021arXiv211206861T}
}

@article{2006:Skilling,
author = {Skilling, John},
year = {2006},
month = {12},
pages = {833-860},
title = {Skilling, J.: Nested sampling for general Bayesian computation. Bayesian Anal. 1(4), 833-860},
volume = {1},
journal = {Bayesian Analysis},
doi = {10.1214/06-BA127}
}

@ARTICLE{2025:O4_methods,
       author = {{The LIGO Scientific Collaboration} and {the Virgo Collaboration} and {the KAGRA Collaboration} and {Abac}, A.~G. and {Abouelfettouh}, I. and {Acernese}, F. and {Ackley}, K. and {Adhicary}, S. and {Adhikari}, D. and {Adhikari}, N. and {Adhikari}, R.~X. and {Adkins}, V.~K. and {Afroz}, S. and {Agarwal}, D. and {Agathos}, M. and {Aghaei Abchouyeh}, M. and {Aguiar}, O.~D. and {Ahmadzadeh}, S. and {Aiello}, L. and {Ain}, A. and {Ajith}, P. and {Akcay}, S. and {Akutsu}, T. and {Albanesi}, S. and {Alfaidi}, R.~A. and {Al-Jodah}, A. and {All{\'e}n{\'e}}, C. and {Allocca}, A. and {Al-Shammari}, S. and {Altin}, P.~A. and {Alvarez-Lopez}, S. and {Amarasinghe}, O. and {Amato}, A. and {Amra}, C. and {Ananyeva}, A. and {Anderson}, S.~B. and {Anderson}, W.~G. and {Andia}, M. and {Ando}, M. and {Andrade}, T. and {Andr{\'e}s-Carcasona}, M. and {Andri{\'c}}, T. and {Anglin}, J. and {Ansoldi}, S. and {Antelis}, J.~M. and {Antier}, S. and {Aoumi}, M. and {Appavuravther}, E.~Z. and {Appert}, S. and {Apple}, S.~K. and {Arai}, K. and {Araya}, A. and {Araya}, M.~C. and {Arca Sedda}, M. and {Areeda}, J.~S. and {Argianas}, L. and {Aritomi}, N. and {Armato}, F. and {Armstrong}, S. and {Arnaud}, N. and {Arogeti}, M. and {Aronson}, S.~M. and {Ashton}, G. and {Aso}, Y. and {Assiduo}, M. and {Assis de Souza Melo}, S. and {Aston}, S.~M. and {Astone}, P. and {Attadio}, F. and {Aubin}, F. and {AultONeal}, K. and {Avallone}, G. and {Babak}, S. and {Badaracco}, F. and {Badger}, C. and {Bae}, S. and {Bagnasco}, S. and {Bagui}, E. and {Baiotti}, L. and {Bajpai}, R. and {Baka}, T. and {Baker}, T. and {Ball}, M. and {Ballardin}, G. and {Ballmer}, S.~W. and {Banagiri}, S. and {Banerjee}, B. and {Bankar}, D. and {Baptiste}, T.~M. and {Baral}, P. and {Barayoga}, J.~C. and {Barish}, B.~C. and {Barker}, D. and {Barman}, N. and {Barneo}, P. and {Barone}, F. and {Barr}, B. and {Barsotti}, L. and {Barsuglia}, M. and {Barta}, D. and {Bartoletti}, A.~M. and {Barton}, M.~A. and {Bartos}, I. and {Basak}, S. and {Basalaev}, A. and {Bassiri}, R. and {Basti}, A. and {Bates}, D.~E. and {Bawaj}, M. and {Baxi}, P. and {Bayley}, J.~C. and {Baylor}, A.~C. and {Baynard}, II, P.~A. and {Bazzan}, M. and {Bedakihale}, V.~M. and {Beirnaert}, F. and {Bejger}, M. and {Belardinelli}, D. and {Bell}, A.~S. and {Bellie}, D.~S. and {Bellizzi}, L. and {Benoit}, W. and {Bentara}, I. and {Bentley}, J.~D. and {Ben Yaala}, M. and {Bera}, S. and {Bergamin}, F. and {Berger}, B.~K. and {Bernuzzi}, S. and {Beroiz}, M. and {Berry}, C.~P.~L. and {Bersanetti}, D. and {Bertolini}, A. and {Betzwieser}, J. and {Beveridge}, D. and {Bevilacqua}, G. and {Bevins}, N. and {Bhandare}, R. and {Bhat}, S.~A. and {Bhatt}, R. and {Bhattacharjee}, D. and {Bhaumik}, S. and {Bhowmick}, S. and {Biancalana}, V. and {Bianchi}, A. and {Bilenko}, I.~A. and {Billingsley}, G. and {Binetti}, A. and {Bini}, S. and {Binu}, C. and {Birnholtz}, O. and {Biscoveanu}, S. and {Bisht}, A. and {Bitossi}, M. and {Bizouard}, M. -A. and {Blaber}, S. and {Blackburn}, J.~K. and {Blagg}, L.~A. and {Blair}, C.~D. and {Blair}, D.~G. and {Bobba}, F. and {Bode}, N. and {Boileau}, G. and {Boldrini}, M. and {Bolingbroke}, G.~N. and {Bolliand}, A. and {Bonavena}, L.~D. and {Bondarescu}, R. and {Bondu}, F. and {Bonilla}, E. and {Bonilla}, M.~S. and {Bonino}, A. and {Bonnand}, R. and {Booker}, P. and {Borchers}, A. and {Borhanian}, S. and {Boschi}, V. and {Bose}, S. and {Bossilkov}, V. and {Boudon}, A. and {Bozzi}, A. and {Bradaschia}, C. and {Brady}, P.~R. and {Branch}, A. and {Branchesi}, M. and {Braun}, I. and {Briant}, T. and {Brillet}, A. and {Brinkmann}, M. and {Brockill}, P. and {Brockmueller}, E. and {Brooks}, A.~F. and {Brown}, B.~C. and {Brown}, D.~D. and {Brozzetti}, M.~L. and {Brunett}, S. and {Bruno}, G. and {Bruntz}, R. and {Bryant}, J. and {Bu}, Y.},
        title = "{GWTC-4.0: Methods for Identifying and Characterizing Gravitational-wave Transients}",
      journal = {arXiv e-prints},
         year = 2025,
        month = aug,
          eid = {arXiv:2508.18081},
        pages = {arXiv:2508.18081},
          doi = {10.48550/arXiv.2508.18081},
archivePrefix = {arXiv},
       eprint = {2508.18081},
 primaryClass = {gr-qc},
       adsurl = {https://ui.adsabs.harvard.edu/abs/2025arXiv250818081T}
}

@article{2025:Williams_Validating_SMC,
    author = "Williams, Michael J. and Karamanis, Minas and Luo, Yilin and Seljak, Uro{\v{s}}",
    title = "{Validating Sequential Monte Carlo for Gravitational-Wave Inference}",
    eprint = "2506.18977",
    archivePrefix = "arXiv",
    primaryClass = "astro-ph.IM",
    reportNumber = "LIGO-P2500231",
    doi = "10.1093/mnras/staf1458",
    journal = "Mon. Not. Roy. Astron. Soc.",
    volume = "1479",
    pages = "1493",
    year = "2025"
}

@article{2022:Karamanis_Preconditioned,
    author = "Karamanis, Minas and Beutler, Florian and Peacock, John A. and Nabergoj, David and Seljak, Uros",
    title = "{Accelerating astronomical and cosmological inference with preconditioned Monte Carlo}",
    eprint = "2207.05652",
    archivePrefix = "arXiv",
    primaryClass = "astro-ph.IM",
    doi = "10.1093/mnras/stac2272",
    journal = "Mon. Not. Roy. Astron. Soc.",
    volume = "516",
    number = "2",
    pages = "1644--1653",
    year = "2022"
}

@article{2006:Doucet_SMC,
    author = {Del Moral, Pierre and Doucet, Arnaud and Jasra, Ajay},
    title = {Sequential Monte Carlo Samplers},
    journal = {J. R. Stat. Soc. Ser. B Stat. Methodol},
    volume = {68},
    number = {3},
    pages = {411-436},
    year = {2006},
    month = {05},
    issn = {1369-7412},
    doi = {10.1111/j.1467-9868.2006.00553.x},
    url = {https://doi.org/10.1111/j.1467-9868.2006.00553.x},
    eprint = {https://academic.oup.com/jrsssb/article-pdf/68/3/411/49795343/jrsssb_68_3_411.pdf},


}

@article{2021:Williams_Nessai,
    author = "Williams, Michael J. and Veitch, John and Messenger, Chris",
    title = "{Nested sampling with normalizing flows for gravitational-wave inference}",
    eprint = "2102.11056",
    archivePrefix = "arXiv",
    primaryClass = "gr-qc",
    doi = "10.1103/PhysRevD.103.103006",
    journal = "Phys. Rev. D",
    volume = "103",
    number = "10",
    pages = "103006",
    year = "2021"
}

@article{2023:Wong_JIM,
    author = "Wong, Kaze W. K. and Isi, Maximiliano and Edwards, Thomas D. P.",
    title = "{Fast Gravitational-wave Parameter Estimation without Compromises}",
    eprint = "2302.05333",
    archivePrefix = "arXiv",
    primaryClass = "astro-ph.IM",
    doi = "10.3847/1538-4357/acf5cd",
    journal = "Astrophys. J.",
    volume = "958",
    number = "2",
    pages = "129",
    year = "2023"
}

@article{2021:Dax_DINGO,
    author = {Dax, Maximilian and Green, Stephen R. and Gair, Jonathan and Macke, Jakob H. and Buonanno, Alessandra and Sch{\"o}lkopf, Bernhard},
    title = "{Real-Time Gravitational Wave Science with Neural Posterior Estimation}",
    eprint = "2106.12594",
    archivePrefix = "arXiv",
    primaryClass = "gr-qc",
    reportNumber = "LIGO-P2100223",
    doi = "10.1103/PhysRevLett.127.241103",
    journal = "Phys. Rev. Lett.",
    volume = "127",
    number = "24",
    pages = "241103",
    year = "2021"
}

@INCOLLECTION{2011:Neal_HMC,
       author = {{Neal}, Radford},
        title = "{MCMC Using Hamiltonian Dynamics}",
    booktitle = {Handbook of Markov Chain Monte Carlo},
         year = 2011,
        pages = {113-162},
          doi = {10.1201/b10905},
       adsurl = {https://ui.adsabs.harvard.edu/abs/2011hmcm.book..113N}
}

@ARTICLE{2019:Thrane_Talbot,
       author = {{Thrane}, Eric and {Talbot}, Colm},
        title = "{An introduction to Bayesian inference in gravitational-wave astronomy: Parameter estimation, model selection, and hierarchical models}",
      journal = {"Publications of the Astronomical Society of Australia"},
         year = 2019,
        month = mar,
       volume = {36},
          eid = {e010},
        pages = {e010},
          doi = {10.1017/pasa.2019.2},
archivePrefix = {arXiv},
       eprint = {1809.02293},
 primaryClass = {astro-ph.IM},
       adsurl = {https://ui.adsabs.harvard.edu/abs/2019PASA...36...10T}
}

@article{hinne2025introduction,
  title={An introduction to Sequential Monte Carlo for Bayesian inference and model comparison—with examples for psychology and behavioral science},
  author={Hinne, Max},
  journal={Behavior Research Methods},
  volume={57},
  number={5},
  pages={125},
  year={2025},
  publisher={Springer}
}

@ARTICLE{2015:LAL_inference,
       author = {{Veitch}, J. and {Raymond}, V. and {Farr}, B. and {Farr}, W. and {Graff}, P. and {Vitale}, S. and {Aylott}, B. and {Blackburn}, K. and {Christensen}, N. and {Coughlin}, M. and {Del Pozzo}, W. and {Feroz}, F. and {Gair}, J. and {Haster}, C.-J. and {Kalogera}, V. and {Littenberg}, T. and {Mandel}, I. and {O'Shaughnessy}, R. and {Pitkin}, M. and {Rodriguez}, C. and {R{\"o}ver}, C. and {Sidery}, T. and {Smith}, R. and {Van Der Sluys}, M. and {Vecchio}, A. and {Vousden}, W. and {Wade}, L.},
        title = "{Parameter estimation for compact binaries with ground-based gravitational-wave observations using the LALInference software library}",
      journal = {Phys. Rev. D},
         year = 2015,
        month = feb,
       volume = {91},
       number = {4},
          eid = {042003},
        pages = {042003},
          doi = {10.1103/PhysRevD.91.042003},
archivePrefix = {arXiv},
       eprint = {1409.7215},
 primaryClass = {gr-qc},
       adsurl = {https://ui.adsabs.harvard.edu/abs/2015PhRvD..91d2003V}
}

@article{2024:De_Santi,
    author = "De Santi, Federico and Razzano, Massimiliano and Fidecaro, Francesco and Muccillo, Luca and Papalini, Lucia and Patricelli, Barbara",
    title = "{Deep learning to detect gravitational waves from binary close encounters: Fast parameter estimation using normalizing flows}",
    eprint = "2404.12028",
    archivePrefix = "arXiv",
    primaryClass = "gr-qc",
    doi = "10.1103/PhysRevD.109.102004",
    journal = "Phys. Rev. D",
    volume = "109",
    number = "10",
    pages = "102004",
    year = "2024"
}

@article{2023:Relative_Binning,
    author = "Krishna, Kruthi and Vijaykumar, Aditya and Ganguly, Apratim and Talbot, Colm and Biscoveanu, Sylvia and George, Richard N. and Williams, Natalie and Zimmerman, Aaron",
    title = "{Accelerated parameter estimation in Bilby with relative binning}",
    eprint = "2312.06009",
    archivePrefix = "arXiv",
    primaryClass = "gr-qc",
    month = "12",
    year = "2023"
}

@article{2020:ROQ,
    author = "Morisaki, Soichiro and Raymond, Vivien",
    title = "{Rapid Parameter Estimation of Gravitational Waves from Binary Neutron Star Coalescence using Focused Reduced Order Quadrature}",
    eprint = "2007.09108",
    archivePrefix = "arXiv",
    primaryClass = "gr-qc",
    doi = "10.1103/PhysRevD.102.104020",
    journal = "Phys. Rev. D",
    volume = "102",
    number = "10",
    pages = "104020",
    year = "2020"
}

@article{2025:DEEP_HMC,
    author = "Perret, Jules and Ar{\'e}ne, Marc and Porter, Edward K.",
    title = "{DeepHMC : a deep-neural-network acclerated Hamiltonian Monte Carlo algorithm for binary neutron star parameter estimation}",
    eprint = "2505.02589",
    archivePrefix = "arXiv",
    primaryClass = "gr-qc",
    month = "5",
    year = "2025"
}

@article{2022:Cogwheel,
    author = "Roulet, Javier and Olsen, Seth and Mushkin, Jonathan and Islam, Tousif and Venumadhav, Tejaswi and Zackay, Barak and Zaldarriaga, Matias",
    title = "{Removing degeneracy and multimodality in gravitational wave source parameters}",
    eprint = "2207.03508",
    archivePrefix = "arXiv",
    primaryClass = "gr-qc",
    doi = "10.1103/PhysRevD.106.123015",
    journal = "Phys. Rev. D",
    volume = "106",
    number = "12",
    pages = "123015",
    year = "2022"
}

@incollection{2001:SMC_introduction,
  title={An introduction to sequential Monte Carlo methods},
  author={Doucet, Arnaud and De Freitas, Nando and Gordon, Neil},
  booktitle={Sequential Monte Carlo methods in practice},
  pages={3--14},
  year={2001},
  publisher={Springer}
}

@article{2022:SMC_invitation,
  title={An invitation to sequential Monte Carlo samplers},
  author={Dai, Chenguang and Heng, Jeremy and Jacob, Pierre E and Whiteley, Nick},
  journal={Journal of the American Statistical Association},
  volume={117},
  number={539},
  pages={1587--1600},
  year={2022},
  publisher={Taylor \& Francis}
}

@ARTICLE{2020:Dynesty,
       author = {{Speagle}, Joshua S.},
        title = "{DYNESTY: a dynamic nested sampling package for estimating Bayesian posteriors and evidences}",
      journal = {"Mon. Not. Roy. Astron. Soc."},
         year = 2020,
        month = apr,
       volume = {493},
       number = {3},
        pages = {3132-3158},
          doi = {10.1093/mnras/staa278},
archivePrefix = {arXiv},
       eprint = {1904.02180},
 primaryClass = {astro-ph.IM},
       adsurl = {https://ui.adsabs.harvard.edu/abs/2020MNRAS.493.3132S}
}

@article{2019:bilby_paper,
    author = "Ashton, Gregory and others",
    title = "{BILBY: A user-friendly Bayesian inference library for gravitational-wave astronomy}",
    eprint = "1811.02042",
    archivePrefix = "arXiv",
    primaryClass = "astro-ph.IM",
    doi = "10.3847/1538-4365/ab06fc",
    journal = "Astrophys. J. Suppl.",
    volume = "241",
    number = "2",
    pages = "27",
    year = "2019"
}

@ARTICLE{2011:NUTS,
       author = {{Hoffman}, Matthew D. and {Gelman}, Andrew},
        title = "{The No-U-Turn Sampler: Adaptively Setting Path Lengths in Hamiltonian Monte Carlo}",
      journal = {arXiv e-prints},
         year = 2011,
        month = nov,
          eid = {arXiv:1111.4246},
        pages = {arXiv:1111.4246},
          doi = {10.48550/arXiv.1111.4246},
archivePrefix = {arXiv},
       eprint = {1111.4246},
 primaryClass = {stat.CO},
       adsurl = {https://ui.adsabs.harvard.edu/abs/2011arXiv1111.4246H}
}

@article{2021:Multibanding_MOrisaki,
    author = "Morisaki, Soichiro",
    title = "{Accelerating parameter estimation of gravitational waves from compact binary coalescence using adaptive frequency resolutions}",
    eprint = "2104.07813",
    archivePrefix = "arXiv",
    primaryClass = "gr-qc",
    doi = "10.1103/PhysRevD.104.044062",
    journal = "Phys. Rev. D",
    volume = "104",
    number = "4",
    pages = "044062",
    year = "2021"
}

@article{2011:Riemann_Girolami,
  title={Riemann manifold langevin and hamiltonian monte carlo methods},
  author={Girolami, Mark and Calderhead, Ben},
  journal={J. R. Stat. Soc. Ser. B Stat. Methodol},
  volume={73},
  number={2},
  pages={123--214},
  year={2011},
  publisher={Oxford University Press}
}

@ARTICLE{2024:Blackjax,
       author = {{Cabezas}, Alberto and {Corenflos}, Adrien and {Lao}, Junpeng and {Louf}, R{\'e}mi and {Carnec}, Antoine and {Chaudhari}, Kaustubh and {Cohn-Gordon}, Reuben and {Coullon}, Jeremie and {Deng}, Wei and {Duffield}, Sam and {Dur{\'a}n-Mart{\'\i}n}, Gerardo and {Elantkowski}, Marcin and {Foreman-Mackey}, Dan and {Gregori}, Michele and {Iguaran}, Carlos and {Kumar}, Ravin and {Lysy}, Martin and {Murphy}, Kevin and {Orduz}, Juan Camilo and {Patel}, Karm and {Wang}, Xi and {Zinkov}, Rob},
        title = "{BlackJAX: Composable Bayesian inference in JAX}",
      journal = {arXiv e-prints},
         year = 2024,
        month = feb,
          eid = {arXiv:2402.10797},
        pages = {arXiv:2402.10797},
          doi = {10.48550/arXiv.2402.10797},
archivePrefix = {arXiv},
       eprint = {2402.10797},
 primaryClass = {cs.MS},


}

@article{2022:Ashton_NS,
    author = "Ashton, Greg and others",
    title = "{Nested sampling for physical scientists}",
    eprint = "2205.15570",
    archivePrefix = "arXiv",
    primaryClass = "stat.CO",
    doi = "10.1038/s43586-022-00121-x",
    journal = "Nature",
    volume = "2",
    year = "2022"
}

@article{2025:ASPIRE,
    author = "Williams, Michael J.",
    title = "{Accelerated Sequential Posterior Inference via Reuse for Gravitational-Wave Analyses}",
    eprint = "2511.04218",
    archivePrefix = "arXiv",
    primaryClass = "hep-ex",
    reportNumber = "LIGO-P2500623",
    month = "11",
    year = "2025"
}

@article{2025:full_hier_mancarella,
    author = "Mancarella, Michele and Gerosa, Davide",
    title = "{Sampling the full hierarchical population posterior distribution in gravitational-wave astronomy}",
    eprint = "2502.12156",
    archivePrefix = "arXiv",
    primaryClass = "gr-qc",
    doi = "10.1103/PhysRevD.111.103012",
    journal = "Phys. Rev. D",
    volume = "111",
    number = "10",
    pages = "103012",
    year = "2025"
}

@ARTICLE{2024_Persistent,
       author = {{Karamanis}, Minas and {Seljak}, Uro{\v{s}}},
        title = "{Persistent Sampling: Enhancing the Efficiency of Sequential Monte Carlo}",
      journal = {arXiv e-prints},
         year = 2024,
          eid = {arXiv:2407.20722},
        pages = {arXiv:2407.20722},
          doi = {10.48550/arXiv.2407.20722},
archivePrefix = {arXiv},
       eprint = {2407.20722},
 primaryClass = {stat.ML},
       adsurl = {https://ui.adsabs.harvard.edu/abs/2024arXiv240720722K}
}

@article{2023:Ripple,
    author = "Edwards, Thomas D. P. and Wong, Kaze W. K. and Lam, Kelvin K. H. and Coogan, Adam and Foreman-Mackey, Daniel and Isi, Maximiliano and Zimmerman, Aaron",
    title = "{Differentiable and hardware-accelerated waveforms for gravitational wave data analysis}",
    eprint = "2302.05329",
    archivePrefix = "arXiv",
    primaryClass = "astro-ph.IM",
    doi = "10.1103/PhysRevD.110.064028",
    journal = "Phys. Rev. D",
    volume = "110",
    number = "6",
    pages = "064028",
    year = "2024"
}

@article{2020:bilby_ROmero,
    author = "Romero-Shaw, I. M. and others",
    title = "{Bayesian inference for compact binary coalescences with bilby: validation and application to the first LIGO{\textendash}Virgo gravitational-wave transient catalogue}",
    eprint = "2006.00714",
    archivePrefix = "arXiv",
    primaryClass = "astro-ph.IM",
    doi = "10.1093/mnras/staa2850",
    journal = "Mon. Not. Roy. Astron. Soc.",
    volume = "499",
    number = "3",
    pages = "3295--3319",
    year = "2020"
}

@article{2024:Robust_Wouters,
    author = "Wouters, Thibeau and Pang, Peter T. H. and Dietrich, Tim and Van Den Broeck, Chris",
    title = "{Robust parameter estimation within minutes on gravitational wave signals from binary neutron star inspirals}",
    eprint = "2404.11397",
    archivePrefix = "arXiv",
    primaryClass = "astro-ph.IM",
    doi = "10.1103/PhysRevD.110.083033",
    journal = "Phys. Rev. D",
    volume = "110",
    number = "8",
    pages = "083033",
    year = "2024"
}

@INCOLLECTION{2011:HMC_NEAL,
       author = {{Neal}, Radford},
        title = "{MCMC Using Hamiltonian Dynamics}",
    booktitle = {Handbook of Markov Chain Monte Carlo},
         year = 2011,
        pages = {113-162},
          doi = {10.1201/b10905},
       adsurl = {https://ui.adsabs.harvard.edu/abs/2011hmcm.book..113N}
}

@article{2018:ashton_bilby,
    author = "Ashton, Gregory and others",
    title = "{BILBY: A user-friendly Bayesian inference library for gravitational-wave astronomy}",
    eprint = "1811.02042",
    archivePrefix = "arXiv",
    primaryClass = "astro-ph.IM",
    doi = "10.3847/1538-4365/ab06fc",
    journal = "Astrophys. J. Suppl.",
    volume = "241",
    number = "2",
    pages = "27",
    year = "2019"
}

@article{Gabbard:2019rde,
    author = "Gabbard, Hunter and Messenger, Chris and Heng, Ik Siong and Tonolini, Francesco and Murray-Smith, Roderick",
    title = "{Bayesian parameter estimation using conditional variational autoencoders for gravitational-wave astronomy}",
    eprint = "1909.06296",
    archivePrefix = "arXiv",
    primaryClass = "astro-ph.IM",
    doi = "10.1038/s41567-021-01425-7",
    journal = "Nature Phys.",
    volume = "18",
    number = "1",
    pages = "112--117",
    year = "2022"
}

@article{Chua:2019wwt,
    author = "Chua, Alvin J. K. and Vallisneri, Michele",
    title = "{Learning Bayesian posteriors with neural networks for gravitational-wave inference}",
    eprint = "1909.05966",
    archivePrefix = "arXiv",
    primaryClass = "gr-qc",
    doi = "10.1103/PhysRevLett.124.041102",
    journal = "Phys. Rev. Lett.",
    volume = "124",
    number = "4",
    pages = "041102",
    year = "2020"
}

@article{Smith:2016qas,
    author = {Smith, Rory and Field, Scott E. and Blackburn, Kent and Haster, Carl-Johan and P{\"u}rrer, Michael and Raymond, Vivien and Schmidt, Patricia},
    title = "{Fast and accurate inference on gravitational waves from precessing compact binaries}",
    eprint = "1604.08253",
    archivePrefix = "arXiv",
    primaryClass = "gr-qc",
    reportNumber = "LIGO-DOCUMENT-NUMBER-P1600096, LIGO-P1600096",
    doi = "10.1103/PhysRevD.94.044031",
    journal = "Phys. Rev. D",
    volume = "94",
    number = "4",
    pages = "044031",
    year = "2016"
}

@article{Cornish:2010kf,
    author = "Cornish, Neil J.",
    title = "{Fast Fisher Matrices and Lazy Likelihoods}",
    eprint = "1007.4820",
    archivePrefix = "arXiv",
    primaryClass = "gr-qc",
    month = "7",
    year = "2010"
}

@article{Lange:2018pyp,
    author = "Lange, Jacob and O'Shaughnessy, Richard and Rizzo, Monica",
    title = "{Rapid and accurate parameter inference for coalescing, precessing compact binaries}",
    eprint = "1805.10457",
    archivePrefix = "arXiv",
    primaryClass = "gr-qc",
    reportNumber = "LIGO DCC P1800084, LIGO-DCC-P1800084",
    month = "5",
    year = "2018"
}

@article{Phenom4,
   title={Frequency-domain gravitational waves from nonprecessing black-hole binaries. I. New numerical waveforms and anatomy of the signal},
   volume={93},
   ISSN={2470-0029},
   url={http://dx.doi.org/10.1103/PhysRevD.93.044006},
   DOI={10.1103/physrevd.93.044006},
   number={4},
   journal={Physical Review D},
   publisher={American Physical Society (APS)},
   author={Husa, Sascha and Khan, Sebastian and Hannam, Mark and Pürrer, Michael and Ohme, Frank and Forteza, Xisco Jiménez and Bohé, Alejandro},
   year={2016},
   month=feb }

@article{Phenom5,
   title={Frequency-domain gravitational waves from nonprecessing black-hole binaries. II. A phenomenological model for the advanced detector era},
   volume={93},
   ISSN={2470-0029},
   url={http://dx.doi.org/10.1103/PhysRevD.93.044007},
   DOI={10.1103/physrevd.93.044007},
   number={4},
   journal={Physical Review D},
   publisher={American Physical Society (APS)},
   author={Khan, Sebastian and Husa, Sascha and Hannam, Mark and Ohme, Frank and Pürrer, Michael and Forteza, Xisco Jiménez and Bohé, Alejandro},
   year={2016},
   month=feb }

@article{Salomone2025Unbiased,
  title = {Unbiased and Consistent Nested Sampling via Sequential Monte Carlo},
  author = {Salomone, Robert and South, Leah F. and Drovandi, Christopher C. and Kroese, Dirk P. and Johansen, Adam M.},
  volume = {87},
  number = {4},
  pages = {1221--1238},
  year = {2025},
  doi = {10.1093/jrsssb/qkaf015},
  url = {https://academic.oup.com/jrsssb/article/87/4/1221/8129577}
}

@software{jax2018github,
  author = {James Bradbury},
  title = {{JAX}},
  url = {http://github.com/jax-ml/jax},
  version = {0.3.13},
  year = {2018},
}

@software{colm_talbot_2024_14025488,
  author       = {Colm Talbot},
  title        = {bilby-dev/bilby: v2.3.0},
  month        = nov,
  year         = 2024,
  publisher    = {Zenodo},
  version      = {v2.3.0},
  doi          = {10.5281/zenodo.14025488},
  url          = {https://doi.org/10.5281/zenodo.14025488},
}

@article{Foreman-Mackey2016, doi = {10.21105/joss.00024}, url = {https://doi.org/10.21105/joss.00024}, year = {2016}, publisher = {The Open Journal}, volume = {1}, number = {2}, pages = {24}, author = {Foreman-Mackey, Daniel}, title = {corner.py: Scatterplot matrices in Python}, journal = {Journal of Open Source Software} }

@ARTICLE{matplotib,
  author={Hunter, John D.},
  journal={Computing in Science \& Engineering}, 
  title={Matplotlib: A 2D Graphics Environment}, 
  year={2007},
  volume={9},
  number={3},
  pages={90-95},
  doi={10.1109/MCSE.2007.55}}

@ARTICLE{scipy,
       author = {{Virtanen}, Pauli and {Gommers}, Ralf and {Oliphant}, Travis E. and {Haberland}, Matt and {Reddy}, Tyler and {Cournapeau}, David and {Burovski}, Evgeni and {Peterson}, Pearu and {Weckesser}, Warren and {Bright}, Jonathan and {van der Walt}, St{\'e}fan J. and {Brett}, Matthew and {Wilson}, Joshua and {Millman}, K. Jarrod and {Mayorov}, Nikolay and {Nelson}, Andrew R.~J. and {Jones}, Eric and {Kern}, Robert and {Larson}, Eric and {Carey}, C.~J. and {Polat}, {\.I}lhan and {Feng}, Yu and {Moore}, Eric W. and {VanderPlas}, Jake and {Laxalde}, Denis and {Perktold}, Josef and {Cimrman}, Robert and {Henriksen}, Ian and {Quintero}, E.~A. and {Harris}, Charles R. and {Archibald}, Anne M. and {Ribeiro}, Ant{\^o}nio H. and {Pedregosa}, Fabian and {van Mulbregt}, Paul and {SciPy 1.  0 Contributors}},
        title = "{SciPy 1.0: fundamental algorithms for scientific computing in Python}",
      journal = {Nature Medicine},
         year = 2020,
        month = feb,
       volume = {17},
        pages = {261-272},
          doi = {10.1038/s41592-019-0686-2},
}

@article{numpy,
  title={Array programming with NumPy},
  author={Harris, Charles R and Millman, K Jarrod and Van Der Walt, St{\'e}fan J and Gommers, Ralf and Virtanen, Pauli and Cournapeau, David and Wieser, Eric and Taylor, Julian and Berg, Sebastian and Smith, Nathaniel J and others},
  journal={nature},
  volume={585},
  number={7825},
  pages={357--362},
  year={2020},
  publisher={Nature Publishing Group UK London}
}

@article{pesummary,
title = {PESummary: The code agnostic Parameter Estimation Summary page builder},
journal = {SoftwareX},
volume = {15},
pages = {100765},
year = {2021},
issn = {2352-7110},
doi = {https://doi.org/10.1016/j.softx.2021.100765},
url = {https://www.sciencedirect.com/science/article/pii/S2352711021000856},
author = {Charlie Hoy and Vivien Raymond}
}

@article{150914,
    author = "Abbott, B. P. and others",
    collaboration = "LIGO Scientific, Virgo",
    title = "{Observation of Gravitational Waves from a Binary Black Hole Merger}",
    eprint = "1602.03837",
    archivePrefix = "arXiv",
    primaryClass = "gr-qc",
    reportNumber = "LIGO-P150914",
    doi = "10.1103/PhysRevLett.116.061102",
    journal = "Phys. Rev. Lett.",
    volume = "116",
    number = "6",
    pages = "061102",
    year = "2016"
}

@article{DUANE1987216,
title = {Hybrid Monte Carlo},
journal = {Physics Letters B},
volume = {195},
number = {2},
pages = {216-222},
year = {1987},
issn = {0370-2693},
doi = {https://doi.org/10.1016/0370-2693(87)91197-X},
url = {https://www.sciencedirect.com/science/article/pii/037026938791197X},
author = {Simon Duane and A.D. Kennedy and Brian J. Pendleton and Duncan Roweth}
}
\endgroup

\end{document}